\documentclass{article}

\usepackage{arxiv}

\usepackage[utf8]{inputenc} 
\usepackage[T1]{fontenc}    
\usepackage{hyperref}       
\usepackage{url}            
\usepackage{booktabs}       
\usepackage{amsfonts}       
\usepackage{nicefrac}       
\usepackage{microtype}      
\usepackage{lipsum}
\usepackage{graphicx}
\graphicspath{ {./images/} }

\usepackage{natbib}
\usepackage{placeins}
\usepackage{subfig}
\usepackage{url}
\usepackage{amsmath}
\usepackage{xltabular}
\usepackage{booktabs}
\usepackage{graphicx}
\newtheorem{theorem*}{Theorem}

\title{A Framework for Generating Realistic Synthetic Tabular Data in a Randomized Controlled Trial Setting}

\author{
 Niki Z. Petrakos \\
  Department of Epidemiology Biostatistics and Occupational Health \\
  McGill University \\
  Montreal, Canada \\
  \texttt{niki.petrakos@mail.mcgill.ca} \\
   \And
 Erica E. M. Moodie \\
  Department of Epidemiology Biostatistics and Occupational Health\\
  McGill University\\
  Montreal, Canada \\
  \texttt{erica.moodie@mcgill.ca} \\
  \And
 Nicolas Savy \\
  Département Mathématiques et Informatique \\
  Université Toulouse - Jean Jaures \\
  Toulouse, France \\
  \texttt{nicolas.savy@math.univ-toulouse.fr} \\
}

\begin{document}
\maketitle
\begin{abstract}
Generation of realistic synthetic data has garnered considerable attention in recent years, particularly in the health research domain due to its utility in, for instance, sharing data while protecting patient privacy or determining optimal clinical trial design. While much work has been concentrated on synthetic image generation, generation of realistic and complex synthetic tabular data of the type most commonly encountered in classic epidemiological or clinical studies is still lacking, especially with regards to generating data for randomized controlled trials (RTCs). There is no consensus regarding the best way to generate synthetic tabular RCT data such that the underlying multivariate data distribution is preserved. Motivated by an RCT in the treatment of Human Immunodeficiency Virus, we empirically compared the ability of several strategies and two generation techniques (one machine learning, the other a more classical statistical method) to faithfully reproduce realistic data. Our results suggest that using a sequential generation approach with a R-vine copula model to generate baseline variables, followed by a simple random treatment allocation to mimic the RCT setting, and subsequent regression models for variables post-treatment allocation (such as the trial outcome) is the most effective way to generate synthetic tabular RCT data that capture important and realistic features of the real data.
\end{abstract}


\section{Introduction}\label{introduction_sec1}

Synthetic data generation has become a topic of increased interest in many disciplines, such as finance, climate science, and health research \citep{Assefa2020,Hsu2020,Pezoulas2024}. Across these domains, countless studies require the analysis of complex systems (e.g., testing trading algorithms in hypothetical economic market stressors, the analysis of extreme weather events impacting power grids, or determining the best medical treatment for a group of patients). However, the data required are often difficult to access, usually due to privacy reasons. Even when the data can be accessed, they can be incomplete or otherwise insufficient in answering the scientific question at hand \citep{Figueira2022,Schreck2024}. An increasingly common solution has been to turn towards synthetic data sets that are faithful to the original data and can demonstrate important features of the original data. Though an abundance of synthetic data generation methods exist, many create generated data sets that are much too simplistic and cannot reflect complexities of the real world and the unusual univariate and multivariate distributions that can be observed, especially with mixed data types \citep{Franklin2014,Souli2023,Crowther2013}. In health research in particular, plasmode simulations were introduced in the early 2000s to allow some of the real-world complexity to be captured in the simulated data \citep{Vaughan2009}, but a major drawback of this method is they often focus only on generation of an outcome variable, rather than generating realistic data for a large number of variables at once \citep{Schreck2024}. It may also be the case that repeated simulations are called for, or there may be a need to introduce more variability (such as to protect data privacy \citep{OKeefe2015}). Recently in medical research, where RCTs are often held as the “gold standard” for determining efficacy and safety of various treatments, data generation via "in silico trials" has gained attention. In silico trials are virtual trials that are conducted through computer simulations in order to generate synthetic data such that the generated distribution mimics the original data distribution without simply copying the original observations \citep{Pappalardo2019}. The utility of such trials includes determining, for example, optimal study design of a full-scale RCT, as it would be of great benefit to be able to study different iterations of study designs in a timely, efficient manner to determine which study design to implement in real life \citep{Chen2021, SarramiForoushani2021,Zwep2024}. Of course, as trial designs can be very complex with multiple features to consider (sample size, follow-up time, definition of outcome, etc.), the ability to generate synthetic data that reflect real-world complexities becomes a necessity.

Trials, and indeed a large number of epidemiological and clinical studies, have data that can be represented in tabular form. Tabular data refers to a table of data with each column representing a random variable such that all columns together follow an unknown joint distribution. Each row represents one observation from that (unknown) joint distribution. A major breakthrough in synthetic data generation occurred in 2014, when Goodfellow et al. introduced Generative Adversarial Networks (GANs), a form of deep learning \citep{Goodfellow2014}. GANs have been particularly influential in improving image generation, with many specific GAN architectures designed to handle image data \citep{Mirza2014,Denton2015,Radford2015}. However, considerably less research has been dedicated to the tabular data context (within which RCT data fall), with some notable exceptions of specific variants of GANs (termed GAN architectures) showing promise for this context \citep{Xu2019,Choi2017,Rajabi2022,Walia2020}. Even less work has focused on the context of small, tabular, RCT data in which use cases for synthetic data sets are just beginning to be explored and appreciated \citep{Askin2023,SarramiForoushani2021,Wang2023}, as most work applied to tabular health care data is currently devoted to observational, electronic health records \citep{Hernandez2022}. In comparison to image data, tabular data (and in particular tabular RCT data) have added difficulties for the data generation process, including variables with more complex dependence structures, small sample sizes, variables with different distributions (e.g., continuous and discrete variables), variables with complex distributions (e.g., multi-modal continuous distributions, mixture distributions), and categorical variables with imbalanced class representation. These difficulties are not rare cases for which it is difficult to find examples in real-life applications. Rather, these added complexities are commonplace in tabular RCT data, for which there are often fewer than a few hundred participants (i.e., observations/rows). For example, age and sex are almost always recorded in RCTs, where age usually follows a continuous distribution and sex a discrete distribution. Health scores are another commonly-included variable in RCTs, which often follow a multi-modal continuous distribution. Demographics (such as sex and ethnicity) may have groups that are rarely observed in a given sample (women and people of colour are often not well represented in RCTs) \citep{BodenAlbala2022,Heiat2002}. This leads to imbalanced class representations where the vast majority of observations are in a select few strata. Thus, it is important to consider data generation methods that are able to adequately handle these challenges.  

While machine learning (ML) methods have increased in popularity in the data generation domain, they are not the only methods that hold promise. Some researchers have shown that copula-based generation approaches can also work very well when generating realistic synthetic data \citep{Patki2016,Sun2019,Demeulemeester2023,Zwep2024}. Though copulas were introduced decades ago in the statistics literature \citep{Sklar1959}, their use for synthetic data generation is relatively new. Copulas, and in particular R-vine copulas, are especially attractive for generating tabular RCT data because they can effectively capture univariate distributions by construction, as well as multivariate dependencies between variables (no matter the distribution of the variables).

\begin{figure}
\centering 
\includegraphics[width=0.75\linewidth]{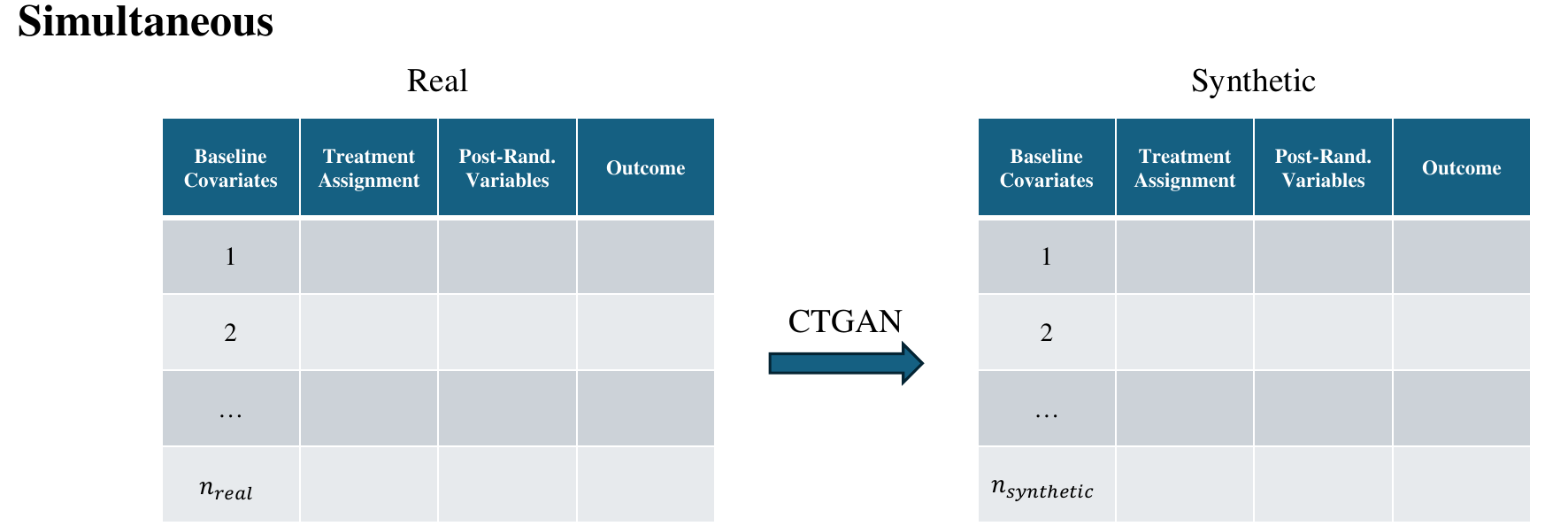}
\includegraphics[width=0.8\linewidth]{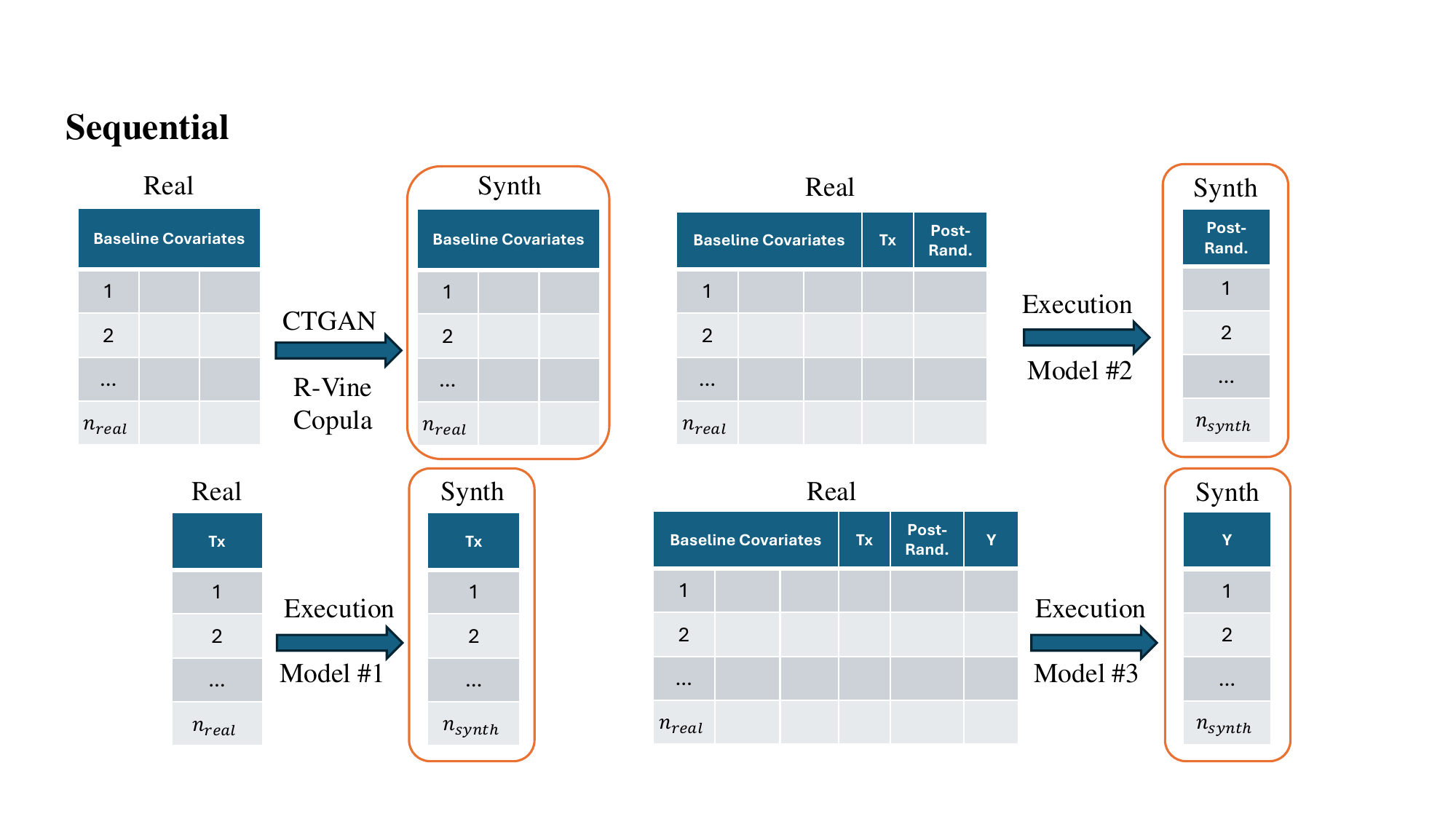}
\caption{Schematic of the steps in the simultaneous versus sequential data generation frameworks. Synth stands for synthetic, Tx stands for treatment assignment, Post-Rand. stands for post treatment randomization variables, and Y stands for the final outcome. $n_{real}$ and $n_{synth}$ represent the number of rows in the real and synthetic data sets, respectively. While $n_{real}$ and $n_{synth}$ need not be equivalent, we set $n_{synth}$ to be equal to $n_{real}$ in the work shown in this paper.}
\label{fig:simultaneousvsequential}
\end{figure}

There currently is no consensus in the literature on how best to generate tabular data for RCTs. Generally, there are two distinct ways of generating tables of data - either simultaneously and often using complex machine learning algorithms, or in a sequential fashion. Sequential data generation has blossomed in a variety of contexts such as music and non-tabular health data \citep{Akbari2018,Dahmen2019}, however it is still rather new in the tabular RCT data domain \citep{Demeulemeester2023,SaintPierre2023}. Figure \ref{fig:simultaneousvsequential} provides a schematic to understanding the difference between the simultaneous and sequential frameworks, where we term "execution models" to be models fit to real data to generate synthetic observations for one variable (often, one outcome) at a time. This sequential approach is derived from the well-known area of agent-based models, which have wide-ranging applications, including public health and infectious disease modeling \citep{Maglio2011,Ainslie2018,Yan2018}. Indeed, these models have also been suggested to determine the design of community-randomized prevention studies \citep{Boren2014}. The basic idea of agent-based modeling is to simulate a baseline population, pre-specify outcomes of interest, simulate changes in the behaviour of the baseline population through predictive modeling, and then measure the outcomes of interest. While much of the machine learning literature provides methodologies for performing data generation in a simultaneous fashion, we hypothesize that a sequential framework, similar to the work flow of agent-based models, would be better suited to the task of generating tabular RCT data. This is due to the inherently sequential nature of RCT data. RCT data have a temporal aspect and causal relationships between variables that may be more naturally captured by sequentially generating data. In an RCT, researchers first collect study participant observations for a set of baseline variables. Then, participants are randomized to treatment. After that, they are often followed for a period of time and follow-up data are collected at certain intervals, perhaps repeatedly or simply once with a final outcome. In a simultaneous data generation framework, such as one that uses GANs, temporal relationships between variables are not considered. Since the algorithm learns the distributions of all variables at once, and generates data for all variables at once, information from variables collected later in time can inform the generation of variables representing earlier time points that would not, in the real world, be influenced by future events or measurements. However, though the simultaneous generation does not mimic how an RCT is carried out in "nature", it is possible that the borrowing of future information to inform variables that are temporally precedent could have the potential to greatly improve the data generation performance when certain variables, no matter the temporal relationship, are highly correlated. 

\FloatBarrier

In this work, we aim to empirically study methods for generating synthetic tabular RCT data such that the real data distribution is preserved and provide recommendations for best practice. To accomplish this aim, we compare the simultaneous framework to the sequential framework. Further, within the sequential framework setting, the performance of a GAN-based algorithm and a copula-based algorithm are compared. This article is organized as follows. In Section \ref{generativealgorithms_sec2}, the generative algorithms compared in this article are described. Section \ref{methods_sec3} details the motivating data set and the simultaneous and sequential frameworks to be compared. Section \ref{results_sec4} presents the results of our empirical studies. Finally, Section \ref{discussion_sec5} concludes with recommendations and a discussion of limitations and future directions for related work.

\section{Generative Algorithms}\label{generativealgorithms_sec2}

Early data generation techniques stem from methods that deal with missing data. While these methods are often thought of as a means to impute missing data, they can also be thought of as data generation techniques (since imputing missing data involves creating values where once there were no values). At first, methods were fairly straightforward and simplistic - mean imputation, or carrying-forward values from past instances. Though easy to implement, these methods suffered from making overly-simplistic assumptions that likely did not represent reality. Then, Synthetic Minority Oversampling Technique (SMOTE) was proposed \citep{Chawla2002}. SMOTE works by selecting a sample from an under-represented stratum for a given discrete variable. Then, it selects one of the sample's $k$ nearest neighbours and interpolates between the selected neighbour and the original sample to generate a synthetic sample. Hence, SMOTE was a popular tool for fixing class imbalances within tabular data, and several extensions were developed \citep{Han2005,Bunkhumpornpat2009,He2008,Douzas2018}. However, these methods were still too simplistic for many contexts, especially those that involved generating data with complex distributions \citep{Wang2023}. Additionally, imputation methods generally create observations by exploiting the data of other individuals in the same data set, whereas the goal of generating an entire synthetic data set (e.g., through generative machine learning algorithms) is to do so by exploiting exogenous data by means of fitted models or learning directly from the data. Hence, the development of GANs made the task of generating data much more realistic and thus added to its popularity. In recent years, another method that has garnered much attention in the machine learning field is Variational Autoencoders, or VAEs \citep{Kingma2013}, which generate synthetic observations through a neural network structure that starts with an encoder, which maps inputted data to a compressed latent space, and ends with a decoder, which maps a sample from the latent space to the original data space and outputs a synthetic observation. For tabular data generation, variants known as Conditional Tabular GAN (CTGAN) and Tabular VAE (TVAE) were proposed to handle the complexities of the tabular data context \citep{Xu2019}. In this work, we did not consider TVAE as it has been shown that they can perform poorly when a minority class of a discrete variable is rarely observed \citep{Kiran2023}, and yet this is a common occurrence in RCT data. We now turn to describing the methods employed in the present comparison.

\subsection{GANs and CTGAN}
\label{subsec:GAN}

A GAN typically consists of two fully-connected neural networks -- a generative model, $G$, and a discriminative model, $D$. The generative model, $G(\boldsymbol{z}; \theta_g): \mathcal{Z} \rightarrow \mathcal{X}$ parameterized by $\theta_g$, takes random noise $\boldsymbol{z} \in \mathcal{Z}$ as input and outputs a vector of synthetic values $G(\boldsymbol{z}) \in \mathcal{X}$, where $\mathcal{X}$ is the real data space. The discriminative model, $D(\boldsymbol{x}; \theta_d): \mathcal{X} \rightarrow [0,1]$ parameterized by $\theta_d$, takes the output from $G$ (the vector of synthetic data) as its input, although $D$ could of course also take real data as an input. Then, $D$ (the discriminative model) outputs a scalar value representing the probability that its input came from the real data. Both models are trained simultaneously in an adversarial fashion, where the generative and discriminative models attempt to satisfy a minimax condition with the following value function, $V(G,D)$:
\begin{equation*}
\label{eq:GANvaluefx}
    \min_{G}\max_{D}V(G,D) = \mathbb{E}_{x \sim p_{X}(x)}[\log D(x)] + \mathbb{E}_{z \sim p_{z}(z)}[\log(1 - D(G(z)))].
\end{equation*}

\noindent
A specific GAN architecture that was proposed by Xu and Veeramachaneni et al. in 2019, CTGAN, aims to handle the additional complexities that come with generating tabular data \citep{Xu2019}. In particular, the authors implemented mode-specific normalization in order to better mimic multi-modal continuous distributions. They also introduced a conditional generator with the idea of training-by-sampling, where observations are generated by conditioning on both a particular discrete variable and one of the observed categories for the given discrete variable, allowing for a more even exploration of all classes of each discrete variable, even for those with unbalanced class distributions. Since CTGANs are the preferred state-of-the-art method for generating tabular data with mixed data types \citep{Kiran2024,Chen2024}, and their implementation is well-documented through the Synthetic Data Vault library in python \citep{Patki2016}, we pursue this approach in our investigations.

\subsection{Copulas and R-Vine Copula}

While complex machine learning models have attracted much attention, there are other statistical methods that also hold potential in generating realistic synthetic data with complex distributions. One such method is copulas \citep{Sklar1959}. A copula is a multidimensional cumulative distribution function (CDF) which relates the marginal distributions of a collection of random variables to the overall joint distribution. A key result that allows for the modeling of a complex joint distribution using copulas is Sklar's Theorem \citep{Sklar1959}, which states that for any joint CDF, there exists a copula representation built on the marginal CDFs of each variable:

\begin{theorem*}[Sklar's Theorem]
\label{sklar_thm1}
If $\mathbf{X}=\{X_1, ..., X_d\}$ is a vector of random variables with joint CDF $F_{X_1,..,X_d}$ and marginal CDFs $F_{X_1},...,F_{X_d}$, then there exists a copula $C$ such that 
\[ F_{X_1,..,X_d}(x_1,...,x_d) = C(F_{X_1}(x_1), ..., F_{X_d}(x_d)). \]
\end{theorem*}
\noindent
Then, applying the chain rule formula, it can be shown that the joint probability density function of $\mathbf{X}$ can be represented as the following decomposition:
\begin{equation*}
\label{eq:copulachainrule}
    f_{X_1,...,X_d}(x_1...,x_d) = c(F_{X_1}(x_1),...,F_{X_d}(x_d)) \times \prod_{i=1}^d f_{X_i}(x_i)
\end{equation*}
where $c()$ is the copula density function. Using recursive conditioning and Bayes formula, the joint marginal density function can also be represented as
\begin{equation*}
        f_{X_1,...,X_d}(x_1...,x_d) = f_{X_d}(x_d) \times f_{X_{d-1}|X_d}(x_{d-1}|x_d) \times f_{X_{d-2}|X_{d-1}, X_d}(x_{d-2}|x_{d-1}, x_d) \times ... \times f_{X_1|X_2,...,X_d}(x_1|x_2,...,x_d)
\end{equation*}
and in general, the conditional marginal distribution can be written as
\begin{equation} 
\label{eq:copuladecomp}
    f_{X_k|\mathbf{Z}}(x_k|\mathbf{z}) = c(F_{X_k|\mathbf{Z}_{-j}}(x_k|\mathbf{z}_{-j}), F_{Z_j|\mathbf{Z}_{-j}}(z_j|\mathbf{z}_{-j})) \times f_{X_k|\mathbf{Z}_{-j}}(x_k|\mathbf{z}_{-j})
\end{equation}
where $\mathbf{Z}$ is a $d$-dimensional vector of random variables and $\mathbf{z}$ is a vector of realizations (observations) of $\mathbf{Z}$, and subscript $-j$ indicates the entire vector excluding variable $j$. Of note, this result shows that each conditional marginal distribution can be written as a product of bi-dimensional copulas and conditional marginal density functions. Hence, this representation is often referred to as the pair-copula construction. Because of this formulation, copulas are very useful in modeling multivariate dependencies. 

By construction, copulas are designed to effectively model univariate distributions, since they utilize the inverse of the empirical distribution of each variable when modeling the marginal CDF. This becomes more clear when describing the data generation process using a copula. For each $X_i, i = 1,...,d$, the first step is to sample $n$ independent observations from a random Uniform, $U$. Then, apply the fitted copula model (fitted to the original data) to the independent $U$ to generate dependent observations, $\Tilde{U}$. Finally, apply the integral transform to map $\Tilde{U}$ to the $X_i$ data space via the (inverse of the) empirical distribution of the observed $X_i$.  

Though copula models date back to 1959, vine copulas (of which R-vine copulas are a subset) were developed in the late 1990s and 2000s \citep{Joe1996, BedfordCooke2002, Aas2009} as a specific type of copula that invokes a nested tree (or "vine") structure, where pairs of copulas are utilized to build the entire copula structure, and hence are useful in modeling the dependence between variables in a high-dimensional, multivariable context \citep{Genest2024}. This is because the decomposition shown in \ref{eq:copuladecomp} is far from unique in our data generation context. The decomposition is unique if all random variables are strictly continuous \citep{Sklar1959}, which is very rarely the case in RCT data. R-vine copulas are useful in a high-dimensional multivariable setting because they allow for a structure to be applied to the non-unique decomposition, and there exists an algorithm to traverse said vine structure to effectively model the joint CDF. Due to this utility in a multivariable context, we chose to utilize specifically R-vine copulas as another method to consider in our data generation framework.

\section{Methods}\label{methods_sec3}

We first describe the data set used in this paper to provide context for the data generation frameworks and illustrate more easily how the different frameworks operate in an RCT data setting.

\subsection{Experimental Tools -- Data Description}

The data used are from the AIDS Clinical Trials Group Study 175, ACTG 175 \citep{Hammer1996}, which compared four HIV therapeautic arms considering mono- or combined treatment among adolescents and adults living with HIV and whose CD4 cell counts were between 200 and 500 per cubic millimeter. 2139 patients were recruited from 43 AIDS Clinical Trials Units and nine National Hemophilia Foundation locations in the United States and Puerto Rico. To be eligible to participate in the study, patients had to satisfy the following criteria: be at least 12 years old with a laboratory-confirmed HIV infection, have a CD4 count between 200 and 500 per cubic millimeter in the 30 days leading up to treatment randomization, have no history of acquired immunodeficiency syndrome (AIDS), and have a Karnofsky score of 70 or higher (which is a health score that measures a patient's functional status, with a maximum score of 100). Study participants were randomly assigned to one of four treatments at baseline (zidovudine only -- this was the baseline comparator, zidovudine and didanosine, zidovudine and zalcitabine, or didanosine only) and were followed for a median time of 143 weeks. Several baseline covariates were measured, including sex, age, weight, race, hemophilia status, whether a patient identified as homosexual, injection drug use status, Karnofsky score, history of prior antiretroviral therapy (ART), whether a patient had symptomatic HIV infection, and CD4 count. Patients had follow-up visits at weeks 2, 4, and 8, and then every 12 weeks, and CD4 count was measured every 12 weeks starting from week 8. CD4 count at week 96 was not recorded for approximately 37\% of participants. The primary end point was the composite event defined by having a CD4 count decline of at least 50\%, an event indicating progression of HIV to AIDS, or death. If no event was observed, then the participant's outcome was deemed to be censored. Hence, the trial outcome was whether the composite event was observed. For a full list of variables and their support, see the Appendix. 

\subsection{Experimental Setup}

To provide a framework to generate realistic data in the RCT context, we compared the frameworks that we term CTGAN Simultaneous versus CTGAN Sequential. We also investigated a framework that we term R-Vine Copula Sequential. Within CTGAN Sequential, we experimented with pre-processing the original data in settings where the original data had bounded or asymmetric distributions. Additionally, we compared three different ways to induce randomness when generating data using sequential regression models. Finally, we considered the degree of complexity or sophistication used in each step of the sequential frameworks. Altogether, we compared seven different data generation frameworks. We explain the setup of each in more detail below.

\subsubsection{CTGAN Simultaneous}

The CTGAN Simultaneous framework involved training a CTGAN model on the entire data set, all at once (i.e., simultaneously). A model was trained on the real data set including all baseline covariates, treatment allocation, post-randomization variables, and the final outcome. Then, this trained model was used to generate observations for all variables, again simultaneously. 

\subsubsection{CTGAN Sequential}

The CTGAN Sequential framework again involved training a CTGAN model on the real data set, but the data generation process proceeded in a sequential fashion such that the natural steps in an RCT were followed and temporal relationships between variables were maintained. The key idea was splitting the real data set into subsets, depending on the time points at which variables were collected in a trial setting. Hence, the first step involved only the baseline variables. First, a CTGAN was trained on the real baseline data, and then the fitted CTGAN was used to generate observations for the baseline variables. Then, what we term "execution models" were fit to each subsequent variable. In a trial setting, the next occurrence was treatment allocation, and since the context involved RCT data, treatment allocation was randomized, independently of any baseline information. The execution model for generating treatment data was simply a probability distribution from which to draw observations. For example, to generate data for the treatment arms in the ACTG 175 trial, samples were drawn from a multinomial distribution with equal probability of observing each of the four treatment arms. Then, following treatment allocation, the post-randomization variables were generated. For example, to generate CD4 count at week 20, an execution model was fit to the real data by regressing CD4 count at week 20 on baseline variables and treatment. Similarly, to generate CD4 count at week 96, another execution model was fit to the real data by regressing CD4 count at week 20 on baseline variables, treatment, and CD4 count at week 20. Generating data for one variable at a time was an easier task than generating data for several variables at once, as modeling and learning the distribution of a single variable was simpler than doing so for a multivariate distribution representing several variables. Hence, statistical regression models were chosen (either linear or logistic) for generating the post randomization variables, which were much more simplistic and thus less resource-intensive compared to a CTGAN. Finally, the last execution model was for generating the outcome variable, and this model was, again, a regression model. For the binary outcome in the ACTG 175 trial, a logistic regression model was fit to generate the outcome, where all other variables from the real data were included as predictors in the model. 

There is an important distinction between \textit{generating} observations and \textit{predicting} observations, which is particularly pertinent when using regression models to generate data. After a regression execution model was fit to the real data, where the response variable in the model was the variable for which we wish to generate data and the predictors in the model were all variables that came before temporally in the trial setting, the next step was to use the fitted model to generate predictions for the response variable (i.e., the variable for which we were generating observations) using the already-generated synthetic data from the baseline CTGAN generated data plus previous execution model-generated data. However, predicting from a simple execution model such as a linear regression lacks the randomness that would be expected in real (or realistic) data. Hence, we adopted an approach familiar from the multiple imputation literature to ensure that any individuals with identical baseline data could still have different values generated for subsequent variables. In this work, we induced randomness in three different ways to compare which was the most effective: \textbf{a)} generated observation = prediction + sample from $N(0, \hat{\sigma}_{resid})$, where $\hat{\sigma}_{resid}$ was the observed standard error of the residuals from the fitted execution model; \textbf{b)} generated observation = prediction + sample from residuals; and \textbf{c)} generated observation was a sample from $\{(\text{prediction} + \text{residuals}) \geq 0 \}$. The third method was implemented by simply rejecting any value of $\text{prediction} + \text{residuals})$ that did not satisfy the bounds of the distribution. In other words, first sample one of the model residuals, then check whether the resulting generated value of $\text{prediction} + \text{residuals})$ was valid. If valid, set the generated value to this sum, and if not, repeat the process. The first two methods of inducing randomness could lead to nonsensical values given the context of the generated variable (e.g., negative values for CD4 count), which motivated the inclusion of the third method where all generated values would be admissible. In our context, inadmissible values were those less than 0, which was why the inequality threshold was set to zero. However, for other contexts, one would simply adjust this threshold value to adapt this strategy to a different context. Additionally, the second and third methods of inducing randomness was included to determine whether omitting the normality assumption from the first method led to better data generation results.

The CTGAN Sequential framework was also implemented with the first method of inducing randomness using pre-processed original data. Certain variables in the real data were transformed before fitting any generation models in the hopes of improving the quality of the generated data. For example, the natural log transformation was applied to CD4 count at baseline, week 20, and week 96, and then the generated data were back-transformed so that the final synthetic data set was in the same scale as the real data. 

Additionally, to deduce the impact of the degree of complexity of models used in the sequential framework, CTGANs were defined for each post-treatment execution model rather than regression models. The purpose of this was to determine whether fitting complex models at each stage in a sequential fit was worth the additional computational power and complexity. When fitting CTGAN models as execution models, sequential steps were still followed. For example, when generating CD4 count at week 20 using a CTGAN, only variables collected up to that point in time were included in the new CTGAN (i.e., baseline covariates, treatment allocation, and CD4 count at week 20), thus replicating some of the work of the baseline CTGAN fit. Then, only the generated data for CD4 count at week 20 was saved and merged with the previously-generated synthetic data (baseline data, treatment). This process was repeated for CD4 count at week 96. Then, to generate data for the outcome, a CTGAN was fit to the entire real data set and only the generated data for the outcome from this CTGAN model was saved and merged with the previously-generated synthetic data. Since the training of a CTGAN involves random noise, there was no need to introduce further randomness when fitting a CTGAN as an execution model (unlike in the regression setting).

\subsubsection{R-Vine Copula Sequential}

The R-Vine Copula Sequential framework was very similar to the CTGAN sequential framework. The main difference, as alluded to in the name, was that the baseline generator was a R-vine copula model fitted to the real baseline data, instead of a CTGAN. However, the execution models were exactly the same regression models as described for the CTGAN sequential framework. For the R-Vine Copula Sequential framework, one version of inducing randomness was implemented -- \textbf{c)} generating observations by sampling from the set of (prediction + residuals) $\geq 0$. Additionally, the original data were not pre-processed before generating synthetic data using this method. The choice to not pre-process data, to pursue only regression for execution models, and to introduce randomness that ensured the synthetic data respected the range of the data distribution were based on preliminary explorations of the methods with the CTGAN approach.

\subsubsection{Simulations}

\begin{figure}[h]
    \centering
    \includegraphics[width=0.75\linewidth]{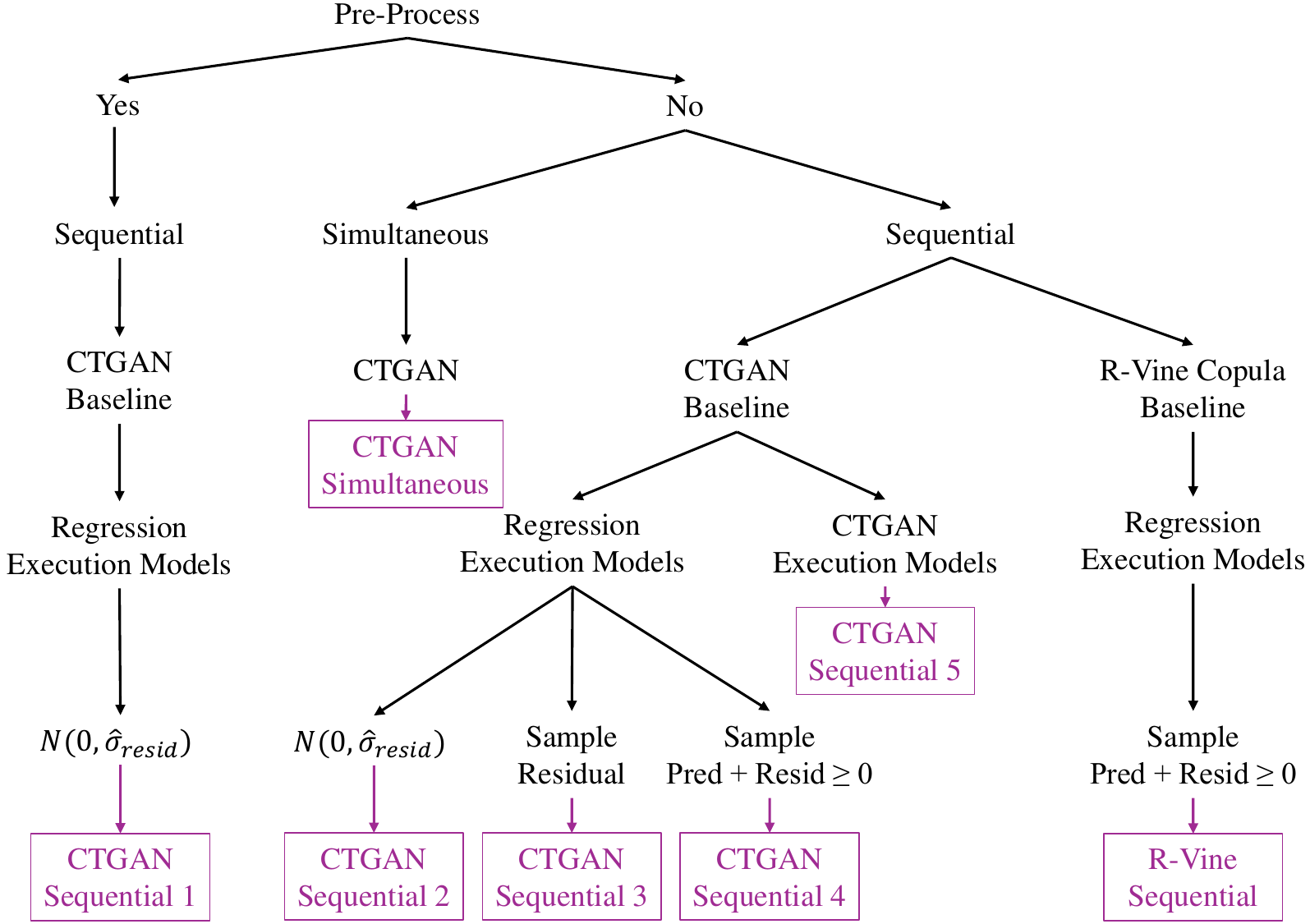}
    \caption{Flowchart of each decision point (data pre-processing, simultaneous versus sequential, CTGAN versus R-vine copula, regression versus CTGAN execution models, method to induce randomness) resulting in the final seven synthetic data generation frameworks.}
    \label{fig:methodflowchart}
\end{figure}

Altogether, we compared seven different data generation frameworks, which we label and characterize as follows. \textbf{CTGAN Simultaneous}: No data pre-processing, CTGAN trained on all data at once; \textbf{CTGAN Sequential 1}: Data pre-processing, CTGAN baseline generator, regression execution models, sample from $N(0, \hat{\sigma}_{resid})$ and add to prediction; \textbf{CTGAN Sequential 2}: No data pre-processing, CTGAN baseline generator, regression execution models, sample from $N(0, \hat{\sigma}_{resid})$ and add to prediction; \textbf{CTGAN Sequential 3}: No data pre-processing, CTGAN baseline generator, regression execution models, sample directly from residuals and add to prediction; \textbf{CTGAN Sequential 4}: No data pre-processing, CTGAN baseline generator, regression execution models, sample from $\{\text{prediction} + \text{residuals} \geq 0\}$; \textbf{CTGAN Sequential 5}: No data pre-processing, CTGAN baseline generator, CTGAN execution models; \textbf{R-Vine Copula Sequential}: No data pre-processing, R-vine copula baseline generator, regression execution models, sample from $\{\text{prediction} + \text{residuals} \geq 0\}$. Figure \ref{fig:methodflowchart} is a flowchart showing key distinctions defined by modeling choices which in turn define each of the seven data generation frameworks.
\FloatBarrier

Since each framework involved a step that induced randomness, the data generation process was repeated 500 times, and performance was compared across all 500 simulation runs. For CTGAN Simultaneous, random noise was incorporated in the fitting of the CTGAN as described in \ref{subsec:GAN}. This was the case for CTGAN Sequential 1-5 as well, though CTGAN Sequential 1-4 also incorporated randomness in the data generation process through the three different methods described earlier for the regression execution models. For R-Vine Sequential, the copula model fit to the original data was also not deterministic due to the non-unique decomposition of the joint distribution, and hence randomness was induced both for the generation of the baseline variables using a R-vine copula, as well as through the incorporation of randomness when generating data using the regression execution models. Therefore, each simulation run involved re-fitting models to generate a new synthetic data set, and then computing metrics for the new synthetic data set. Data generation via CTGANs were conducted in python utilizing the Synthetic Data Vault and SDMetrics libraries,\citep{Patki2016,sdmetrics} and data generation via R-vine copulas were conducted in R utilizing the rvinecopulib package.\citep{rvinecopulib}

\subsubsection{Definition of Models}

Now, we describe in greater detail the models that were fit for each of the seven data generation frameworks. 

In \textbf{CTGAN Simultaneous}, one model was fit in total (a CTGAN), which was trained on the entire data set without any pre-processing of bounded variables.

In \textbf{CTGAN Sequential 1}, five models were fit in total, and the natural log transformation was applied to CD4 count at baseline, week 20, and week 96 in the real data before models were fit. First, a CTGAN was fit to the baseline data only. Then, a multinomial distribution with $n=2139$ and $p = 0.25$ was assumed for the treatment execution model. For CD4 count at week 20, a linear regression model was fit with (the natural log of) CD4 count at week 20 as the response variable and the following variables as the predictors: age, weight, sex, race, hemophilia status, homosexuality identity, intravenous drug use, Karnofsky score, prior non-zidovudine ART usage, zidovudine usage 30 days prior, previous time on ART, ART historical usage, symptomatic HIV status, CD4 count at baseline, and randomized treatment assignment. For CD4 count at week 96, a linear regression model was fit, omitting observations with missing data. CD4 count at week 96 (transformed to be on the natural log scale) was the response variable, and the predictor variables were (the natural log of) CD4 count at week 20 in addition to the same predictor variables as for the CD4 week 20 execution regression model. For the outcome, a logistic regression model was fit, omitting observations with missing data. The response variable was the binary outcome, and the predictor variables were all other variables in the data set (baseline variables, randomized treatment, CD4 count at week 20, and CD4 count at week 96). Finally, the synthetically generated observations for CD4 count at baseline, week 20, and week 96 were exponentiated so that they were on the same scale as the original data.

In \textbf{CTGAN Sequential 2}, similar to CTGAN Sequential 1, a total of five models were fit. The only difference was that there was no data pre-processing.

In \textbf{CTGAN Sequential 3} and \textbf{CTGAN Sequential 4}, same models were fit as in CTGAN Sequential 2 (five models total). Recall that the difference was how randomness was induced when generating values -- CTGAN Sequential 2: sample from $N(0, \hat{\sigma}_{resid})$ and add to prediction; CTGAN Sequential 3: sample directly from residuals and add to prediction, CTGAN Sequential 4: sample from $\{(\text{prediction} + \text{residuals}) \geq 0\}$. That is, in CTGAN Sequential 4, if the randomly selected residual added to the prediction led to an inadmissible value (in this case, a negative CD4 cell count), a new residual was drawn and checked for admissibility.

In \textbf{CTGAN Sequential 5}, again, five models were fit in total, however in this framework, each model except the treatment execution model was a CTGAN. The first CTGAN was fit to only the baseline variables to produce baseline synthetic data. The same multinomial distribution as before was assumed for sampling observations to generate synthetic treatment values. Then, another CTGAN was fit, this time to the baseline variables, treatment, and CD4 count at week 20. Only the generated data for CD4 count at week 20 were saved (the rest of the generated data from this CTGAN were discarded). A third CTGAN was fit to the baseline variables, treatment, CD4 count at week 20, and CD4 count at week 96. Again, only the generated data for CD4 count at week 96 were saved. A final CTGAN was fit to all data (baseline variables, treatment, CD4 count at week 20, CD4 count at week 96, and outcome), and only the generated data for the outcome were saved. 

In \textbf{R-Vine Copula Sequential}, five models were fit in total. First, an R-vine copula model was fit using the baseline data, and then the same execution models as described for CTGAN Sequential 2-4 were fit in this framework as well. As in CTGAN Sequential 4, randomness was incorporated by adding a randomly selected residual that led to an admissible value of the covariate.

\subsection{Experimental Evaluation -- Metrics}

To compare the seven different generation frameworks, a wide range of metrics were used to ensure conclusions were robust to the type of metric used. As was recommended in the literature \citep{Figueira2022}, the quality of the synthetic data (i.e., the closeness of the synthetic data distributions to the original data distributions) was measured by considering univariate and bivariate comparisons. To do so, the simulation was repeated 500 times, where each simulation involved generating a new synthetic data set. For univariate comparisons, the complement of the Kolmogorov Smirnov (KS) statistic was measured for continuous variables, and the complement of the Total Variation Distance (TVD) was measured for discrete variables. The KS statistic is defined as the maximum distance between two empirical CDFs, and hence the complement of the KS statistic represented the "closeness" of the two distributions. The complement of the KS statistic is defined as $1 - \sup_{x \subset X}\left|F^{\text{real}}_{n}(x)-F^{\text{synth}}_{n}(x)\right|$, where $F^{\text{real}}_{n}(x)$ and $F^{\text{synth}}_{n}(x)$ are the empirical CDFs for some continuous variable $X$ with realization $x$ in the real data and synthetic data, respectively. The complement of the TVD is defined as $1 - \frac{1}{2}\sum_{a \subset A}\left|\pi^{\text{real}}_{a}-\pi^{\text{synth}}_{a}\right|$, where $\pi^{\text{real}}_{a}$ and $\pi^{\text{synth}}_{a}$ represent the proportion of stratum $a$ of some discrete variable $A$ in the real and synthetic data, respectively. The TVD represents the difference in proportions for the strata of a given discrete variable between the real and synthetic data. In this context, it can be thought of as the categorical counterpart to the KS statistic. Thus, when utilizing the complement of the KS statistic and the complement of the TVD, higher scores indicated higher quality of the synthetic data generation in terms of fidelity to the real data. 

For bivariate comparisons, a correlation similarity score using the Spearman correlation was computed for pairs of continuous variables and a contingency similarity score was calculated for pairs of discrete variables and discrete and continuous variables. For a pair of variables in which one was discrete and the other continuous, the continuous variable was dichotomized by quartiles. The correlation similarity score is defined as the normalized difference between the real data correlation of a pair of variables and the synthetic data correlation of the same pair of variables: $1 - \frac{1}{2}\left|\rho^{\text{real}}_{XY}-\rho^{\text{synth}}_{XY}\right|$, where $\rho^{\text{real}}_{XY}$ and $\rho^{\text{synth}}_{XY}$ are the Spearman correlation between two continuous variables $X$ and $Y$ in the real and synthetic data sets, respectively. Spearman correlation was calculated rather than Pearson correlation because the former is less parametric. In fact, both were calculated however in our simulations, the results using Spearman versus Pearson correlations were very similar and so choosing one over the other did not make a difference; for brevity, we report only those for the Spearman correlation. The contingency similarity score is defined as the normalized difference between the real data proportions in a contingency table of two variables and the synthetic data proportions in a contingency table of the same variables: $1 - \frac{1}{2}\sum_{a \subset A}\sum_{b \subset B}\left|\pi^{\text{real}}_{ab} - \pi^{\text{synth}}_{ab}\right|$, where $\pi^{\text{real}}_{ab}$ and $\pi^{\text{synth}}_{ab}$ represent the proportion observed in both stratum $a$ of some discrete variable $A$ and stratum $b$ of some discrete variable $B$ in the real and synthetic data, respectively. Again, higher scores indicated higher fidelity of the synthetic to the real data. All univariate and bivariate metrics were computed for all variables and all pairs of variables for each simulation run, and then plotted. The univariate and bivariate distributions were also compared graphically by selecting a single simulation run at random and plotting the synthetic and real data distributions of each variable. Plots for a selection of variables with varying features are presented.

As is often done in the machine learning literature \citep{Xu2019,Zhou2020}, machine learning efficacy metrics were also compared -- precision, recall, and F-1 score. Precision, also known as the positive predictive value in statistics, is the proportion of true positives out of all positives: $\frac{\text{True Positives}}{\text{True Positives + False Positives}}$, where a "positive" is defined as the outcome event having occurred for a given patient. Hence, a "true positive" means the classifier correctly identified that the outcome event occurred for a particular participant, while a "false positive" means the classifier identified that the outcome event occurred for a participant when in truth, the event did not occur. Recall, also known as sensitivity, is the proportion of true positives out of all detected positives by the model: $\frac{\text{True Positives}}{\text{True Positives + False Negatives}}$. The F-1 score is the harmonic mean of precision and recall: $2\cdot\frac{\text{Precision}\cdot\text{Recall}}{\text{Precision} + \text{Recall}}$. Accuracy was not considered because in imbalanced data, it can be a misleading metric \citep{Jeni2013}. First, the real data were split into training and test sets, then a machine learning classifier was trained on the training set and used to predict values for the real data test set. Similarly, the synthetic data were split into training and test sets, and another classifier was trained to the synthetic data training set and then used to predict values for the real data test set. Precision, recall, and the F-1 score were measured for each classifier. Ideally, the scores from the model trained on the real data and from the model trained on the synthetic data should be very similar, thus indicating that the synthetic data were very similar to the real data. Since the choice of classifier could impact the results, two different classifiers were chosen -- XGBoost, which had the ability to deal with missing values, and k-nearest neighbors (KNN), which required a data imputation step \citep{Hastie2001}. When fitting the KNN classifier, $k$ was set to 5 for all simulation runs. Precision, recall, and the F-1 score were measured for each simulation run, and then the differences between the real and synthetic data metrics for precision, recall, and F-1 were plotted across all 500 simulations, for both the XGBoost and KNN classifiers. The plot of absolute differences is shown in Section \ref{results_sec4}, and the plot of relative differences can be found in the Appendix. Values close to zero indicated good performance, meaning the metric value for the classifier trained on real data was similar to the metric value for the classifier trained on synthetic data. Finally, we also measured the total computing time required to generate and evaluate each synthetic data set for each of the seven proposed frameworks.

\section{Results}\label{results_sec4}

\subsection{Graphical Comparisons of the Synthetic and Real Data Distributions}

First, a simulation run was selected at random and the univariate and bivariate distributions were plotted. Though density plots were examined for all variables, only a select few are discussed here for brevity. In all density plots, the pink represents the real data distribution and the blue represents the synthetic data distribution. Each plot shows the distribution generated for each of the seven data generation frameworks, for one simulation run chosen at random. The sequential execution model framework was much more successful at recovering the original data distribution for treatment allocation as compared to the simultaneous framework. This may be due to the nature of the RCT context, where treatment was assigned randomly. When equal probabilities of sampling from each treatment arm were assumed, and then samples were drawn at random to generate the synthetic treatment assignment, the generation process was essentially mimicking the implementation of the real life RCT. In contrast, the simultaneous CTGAN model used all information at once, and thus likely picked up on relationships between treatment assignment and post-treatment randomization variables that then influenced the generation of treatment assignment, even though the treatment assignment was independent of all other variables in the RCT. This was confirmed in the density plot (see Appendix), where the generated treatment values were skewed towards arms 2 and 3.

When comparing the generation performance of each framework for the variable age, for instance, R-Vine Copula Sequential was clearly the most effective at capturing the original data distribution. Pre-processing data (CTGAN Sequential 1) shifted the mean of the synthetic distribution closer to the mean of the real distribution, but the tail of the synthetic distribution exhibited behaviour that did not exist in the real distribution. Also, it was to be expected that CTGAN Sequential 2-5 would have very similar results since they generated the baseline data in the same manner. (The density plots of age for one simulation run across all seven frameworks is included in the Appendix.) Similar results are displayed in Figure \ref{fig:7methods_karnof}, which shows the density plots for the Karnofsky score comparing synthetic to real data. For this continuous variable, which had a multi-modal distribution, CTGAN Simultaneous performed very poorly in mimicking the original data distribution. Again, R-Vine Copula Sequential performed the best, though it of note that the CTGAN Sequential frameworks were able to pick up on the multiple peaks in the original data distributions, even though in these frameworks, the Karnofsky score (a baseline variable) was generated by a CTGAN just like in CTGAN Simultaneous. The difference was that the CTGAN Sequential frameworks trained the CTGAN generative model on only the baseline data, rather than all data simultaneously. This seemed to improve the generation of the baseline variables. In terms of generating the outcome variable (which was binary), most methods performed well, with R-Vine Copula Sequential and CTGAN Sequential 4 showing the best performance and CTGAN Sequential 1 (pre-processing data) showing the worst performance. Refer to the Appendix for the density plots of the generation of the composite binary event outcome.

\begin{figure}[h!]
    \centering
    \includegraphics[width=0.75\linewidth]{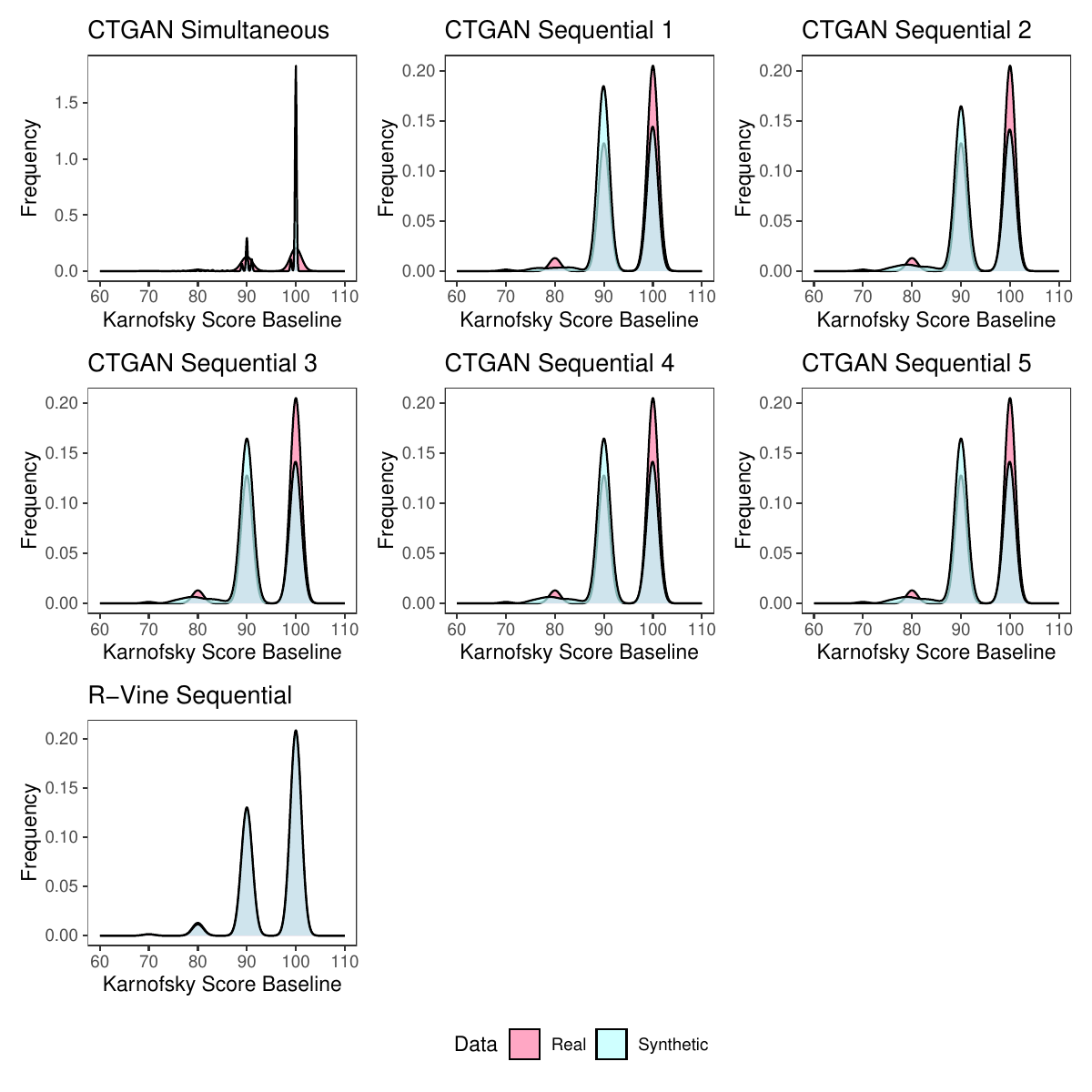}
    \caption{Density plots of Karnofsky score at baseline for a single data generation run as compared to the real data for seven candidate synthetic data generators.}
    \label{fig:7methods_karnof}
\end{figure}

The performance of each method with regards to generating data for variables that change over time was also investigated. In particular, the synthetic and real distributions were plotted for CD4 count at baseline and week 20 (see Appendix), as well as week 96 (Figure \ref{fig:7methods_cd496}). At baseline, R-Vine Copula Sequential was by far the most effective at mimicking the original data distribution. The other frameworks were able to capture the mean of the original distribution, but the spreads were different. At week 20, R-Vine Copula Sequential again outperformed the rest, though CTGAN Sequential 1-4 performed comparably. Both methods that used CTGANs to generate CD4 cell count at week 20 (CTGAN Simultaneous, CTGAN Sequential 5) were much less successful at capturing the real data distribution. At week 96 (where 37\% of participants in the real data had missing values for CD4 count at week 96), CTGAN Simultaneous and CTGAN Sequential 5 performed very poorly at capturing the real data distribution. Again, R-Vine Copula Sequential outperformed the rest, though CTGAN Sequential 3 and 4 showed adequate performance. 

\FloatBarrier
\begin{figure}
    \centering
    \includegraphics[width=0.75\linewidth]{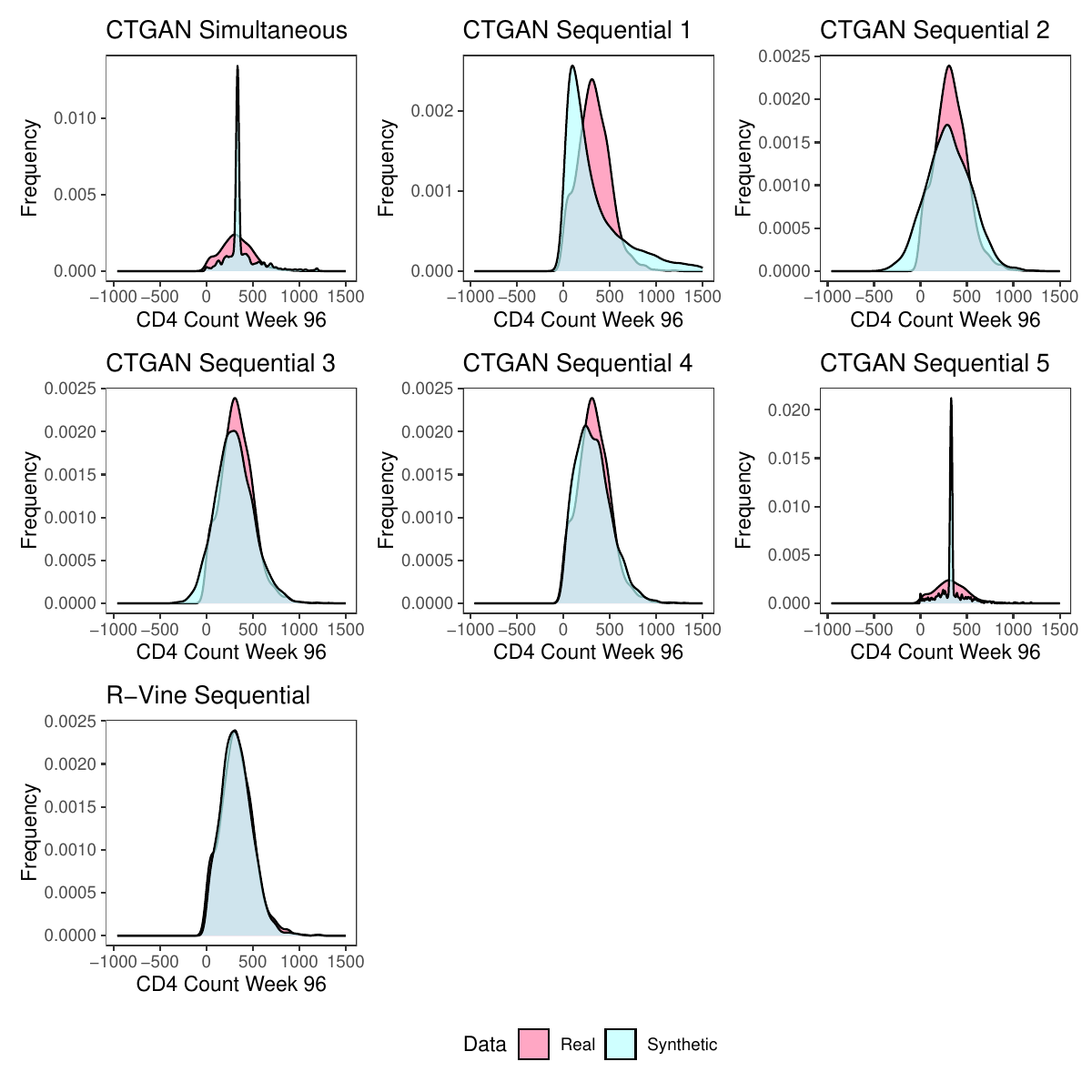}
    \caption{Density plots of CD4 cell count at week 96 for a single data generation run as compared to the real data for seven candidate synthetic data generators.}
    \label{fig:7methods_cd496}
\end{figure}

Additionally, bivariate density plots were employed to evaluate whether relationships between pairs of variables were captured and to verify whether generated data created correlations between variables that were not present in the original data. Several pairs of variables were examined, though two pairs of variables are shown here for brevity: CD4 count at baseline and CD4 count at week 20 (expected to be highly correlated), and age and treatment arm (expected to be uncorrelated). Plots of these bivariate comparisons can be found in the Appendix. R-Vine Copula Sequential was the most successful at capturing the highly-correlated relationship. CTGAN Sequential 4 also showed satisfactory performance, whereas CTGAN Sequential 1 performed the worst. Additionally, CTGAN Sequential 2 and 3 generated inadmissible values for CD4 at week 20 (values less than zero) whereas CTGAN Sequential 4 generated only admissible values. CTGAN Simultaneous and CTGAN Sequential 5 generated distributions that overlapped with the original real-data distribution, but with more noise. All frameworks were successful in not creating relationships between variables when none existed in the real data. R-Vine Copula Sequential performed remarkably well at capturing the real data bivariate distribution for age against treatment arm, where the mean age across treatment arms was very close to the real data, and the spread of age across treatment arms was also very similar to that of the real data. CTGAN Sequential 1 was also moderately successful at mimicking the original bivariate data distribution. The rest of the frameworks performed similarly to one another. 

\subsection{Univariate and Bivariate Performance Metrics}

The seven data generation frameworks were further evaluated based on univariate and bivariate distribution similarity scores. Recall that for univariate continuous distributions, the complement of the KS statistic was employed, and for univariate discrete distributions, the complement of the TVD was calculated. For bivariate distribution comparisons where both variables were continuous, a normalized difference in correlation was utilized. For bivariate distribution comparisons where both variables were discrete or one variable was discrete and the other was continuous, a normalized difference in proportions from the contingency table was used. In all cases, a higher value indicated better performance. 

\begin{figure}[t!]
        \subfloat[Univariate Continuous Similarity Scores]{%
            \includegraphics[width=.48\linewidth]{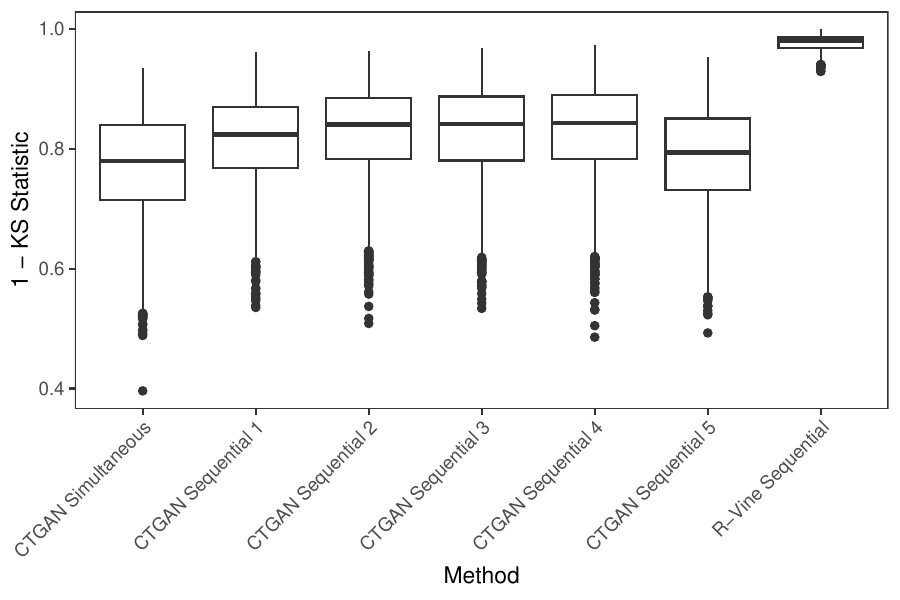}%
            \label{subfig:a}%
        }\hfill
        \subfloat[Univariate Discrete Similarity Scores]{%
            \includegraphics[width=.48\linewidth]{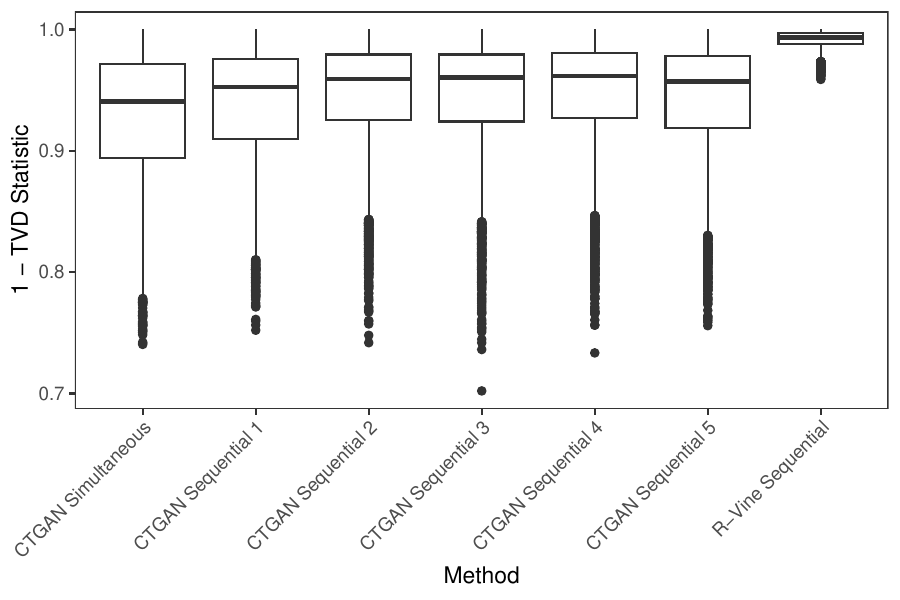}%
            \label{subfig:b}%
        }\\
        \subfloat[Bivariate Continuous Similarity Scores]{%
            \includegraphics[width=.48\linewidth]{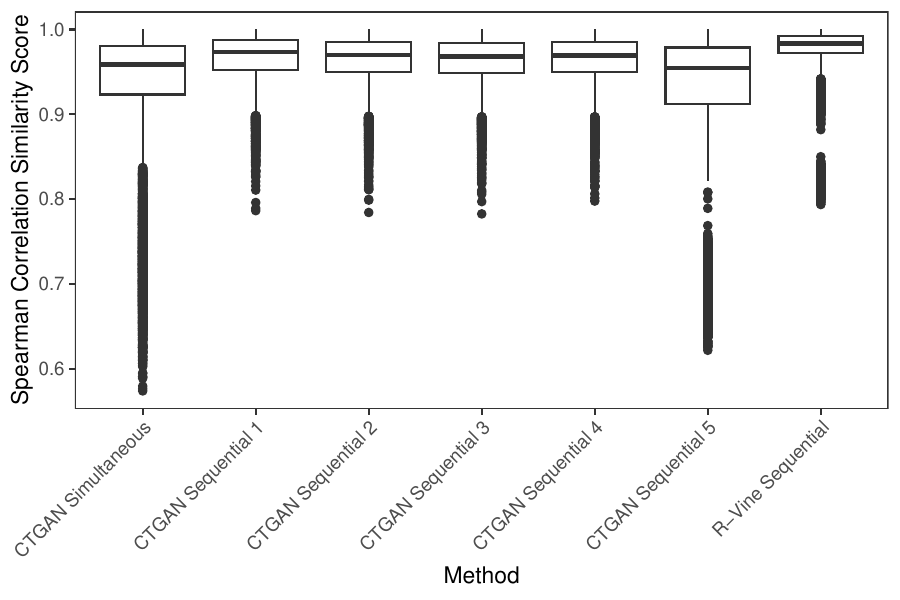}%
            \label{subfig:c}%
        }\hfill
        \subfloat[Bivariate Discrete Similarity Scores]{%
            \includegraphics[width=.48\linewidth]{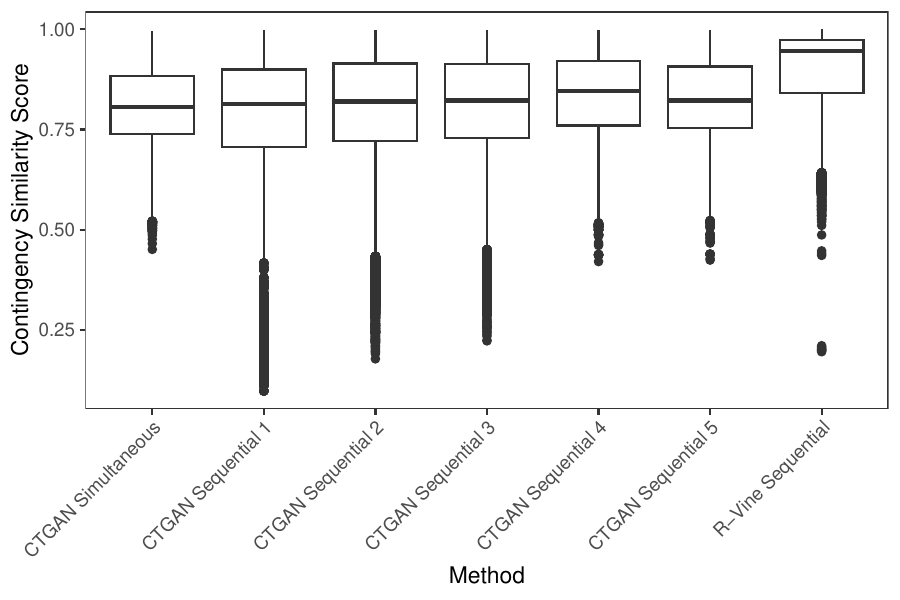}%
            \label{subfig:d}%
        }
        \caption{Comparison of the seven data generation methods by similarity of the synthetic and real distributions. The top left represents univariate similarity scores for continuous variables, and the top right represents univariate similarity scores for discrete variables. The bottom left shows the bivariate correlation similarity scores, and the bottom right shows the bivariate contingency similarity scores. Note that these scores aggregate across multiple variables and across all data generation runs.}
        \label{fig:univarbivarmetrics}
\end{figure}

As shown in the top left of Figure \ref{fig:univarbivarmetrics}, the R-Vine Copula Sequential framework outperformed the other frameworks in terms of capturing univariate continuous distributions in the original data. Not only did it perform better on average, as evidenced by the mean being much higher than the rest, but the spread of values across all 500 simulations was also very tight. This differed from the other frameworks that, at times, performed extremely poorly. The four iterations of CTGAN Sequential utilizing regression models (1-4) all performed similarly (Figure \ref{fig:univarbivarmetrics}, top left panel). When capturing the univariate discrete distributions in the original data, R-Vine Copula Sequential again performed the best (Figure \ref{fig:univarbivarmetrics}, top right panel). Though the difference between R-Vine Copula Sequential and the other frameworks was less extreme when considering the univariate discrete distributions than the univariate continuous distributions, R-Vine Copula Sequential still demonstrated the best performance. This is perhaps not surprising, given the formulation R-vine copulas. In both the univariate continuous and discrete settings, CTGAN Simultaneous had the poorest performance.  

When capturing the correlation between pairs of (continuous) variables, R-Vine Copula Sequential again performed the best, as shown in the bottom left of Figure \ref{fig:univarbivarmetrics}. Though all methods showed adequate performance on average, there were instances in which CTGAN Simultaneous and CTGAN Sequential 5 (the two methods that harnessed CTGANs as the post-treatment execution models) performed very poorly. Again, CTGAN Sequential 1-4 demonstrated very similar results. When comparing the ability to maintain the bivariate relationships between pairs of discrete variables and discrete and continuous variables, there was a clearer distinction between R-Vine Copula Sequential and the CTGAN-based frameworks, as shown in the bottom right of Figure \ref{fig:univarbivarmetrics}. Though there were a few instances when R-Vine Copula Sequential performed sub-optimally, it was again the most successful framework at capturing the bivariate relationships in the real data, on average. 

\begin{figure}[t!]
    \centering
    \includegraphics[width=\linewidth]{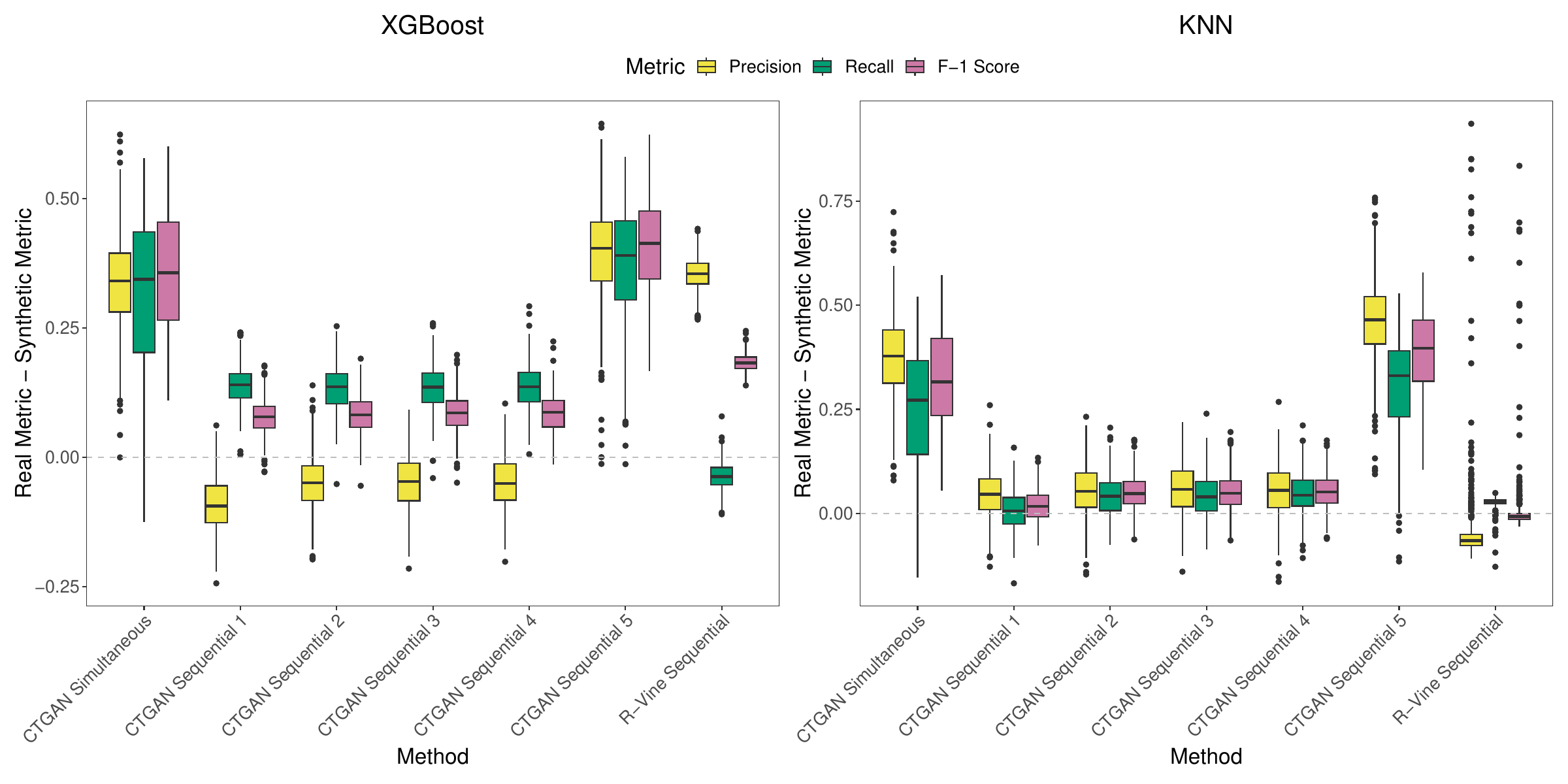}
    \caption{Comparison of the seven data generation methods by similarity of the ML efficacy metrics (precision, recall, F-1 score) for both XGBoost classifiers and KNN classifiers. The vertical axis represents the absolute difference between the real and synthetic metric values. The horizontal dashed line at zero indicates that the metric value for the classifier trained on real data is equal to that of the classifier trained on the synthetic data. A value close to zero means that the two models performed similarly and therefore the synthetic data successfully retained the original data distribution.}
    \label{fig:MLefficacy}
\end{figure}

When interpreting the ML efficacy results, it was less clear which framework performed the best, since the results were highly dependent on the type of classifier (in this case, XGBoost versus KNN), as well as the metric (precision, recall, or F-1 score). One pattern that emerged in Figure \ref{fig:MLefficacy} was that no matter the classifier, CTGAN Simultaneous and CTGAN Sequential 5 performed the worst in generating data with a distribution that closely resembled that of the original data. Also, regardless of the classifier used to assess efficacy, CTGAN Sequential 1-4 performed very similarly, and their results showed rather high performance compared to the rest of the frameworks since the box plots were generally close to zero. This was much more apparent for the KNN classifier. For XGBoost, the R-Vine Sequential method performed well when considering the difference in recall metric values, but performed much worse than CTGAN Sequential 1-4 when considering the difference in precision metric values and the difference in F-1 metric values. For the KNN classifier, while the averages of the difference between metric values for all three metrics for R-Vine Sequential were close to zero, there were some instances in which this framework performed very poorly, and was even the worst out of all seven methods as evidenced by the large outliers. While it is recommended to present these ML efficacy metrics to assess the quality of the synthetic data, it is unfortunately the case that these results were less interpretable compared to the distribution plots and similarity scores. Additionally, it was not clear if the mixed results from the ML efficacy metrics were due to true differences between the real and synthetic data distributions for each framework, or to what extent they were influenced by the choice of classifier or ML metric.

\FloatBarrier

\subsection{Computing Time}

Lastly, we compared the computing time needed for all 500 simulations to generate data and evaluate metrics for each of the seven frameworks. Simulations were performed on a machine with 14core CPU and 16 GB RAM. Ordered from fastest to slowest, the time taken to generate and evaluate synthetic data were the following, in hh:mm:ss format: CTGAN Sequential 2 (07:23:17), CTGAN Sequential 1 (07:56:06), CTGAN Sequential 4 (08:01:45), CTGAN Sequential 3 (08:23:06), CTGAN Simultaneous (10:03:10), R-Vine Sequential (13:32:33), and CTGAN Sequential 5 (32:47:32). All frameworks involving CTGAN were run using Python version 3.9, and the R-Vine Sequential framework was run using R version 4.3.1. R-Vine Sequential took more time than the CTGAN Sequential methods that also utilized regression models, while using CTGAN models everywhere was the most time and resource intensive endeavour. Though these differences in computing time are important, none of the frameworks were so slow as to be prohibitive.

\section{Discussion}\label{discussion_sec5}

In this paper, we developed and compared the performance of seven different data generation frameworks to determine which was the most effective at reproducing the original data distribution of a data set arising from an RCT. We compared the difference between a simultaneous framework and a sequential framework, and within the sequential framework, we evaluated the difference between utilizing a CTGAN versus a R-vine copula to generate the baseline cohort data with various options to address bounded (non-negative) variables. We found that the sequential data generation framework greatly outperformed simultaneous data generation when the task was to generate a synthetic tabular data set in an RCT context that maintained the original data distribution. In particular, the best framework out of all seven that was to not pre-process data, to use an R-vine copula to generate data for baseline variables, fit regression models for post-baseline variables, and induce randomness by drawing a sample from the set of admissible values of prediction plus residuals to generate the synthetic observation. This R-Vine Copula Sequential data generation framework showed great promise, as it outperformed all other frameworks in capturing both the univariate and bivariate distributions in the original data. While its performance for capturing univariate distributions was to be expected, it was surprising to see it outperform CTGAN models when capturing bivariate distributions as well, since both R-vine copula models and CTGANs were designed to capture multivariate dependencies. While complex machine learning methods have gained much popularity and perform very well in certain contexts, our investigations suggest that they underperform for RCT data generation purposes. At first glance, it may appear that generating all data at once using a powerful model such as a CTGAN should be sufficient. We have shown this is not the case -- in an RCT setting, it is necessary to generate data in sequential steps that follow the temporal ordering in a real life RCT. Moreover, using future data to inform the generation of past data negatively impacted performance. 

While these investigations demonstrated many strengths, notably in providing a framework for effectively generating synthetic tabular data in an RCT context, there were also some limitations. While we examined seven possible approaches, we did not consider the fourth permutation when comparing data generation frameworks: simultaneous data generation using an R-vine copula. This latter omission was due to the limitations of the vine structure, where including time-varying treatment and covariates would result in closed loops between variables, which is impossible for a vine. Though our use-case example is a single stage RCT where treatment was not time-varying, we will, in future work, extend these investigations to RCTs where treatments do change over time (such as in Sequential Multiple Assignment Randomized Trials, also known as SMARTs \citep{Collins2007}). Additionally, in this work, the same number of observations were generated as were collected in the original data. We did not explore the potential impact of generating a synthetic data set with more or fewer observations (rows) than the original data set. However, we did not expect that one framework or method would perform better for different generated sample sizes, and we did not believe that exploring the effect of changing the sample size of the synthetic data set would be a scientific question of interest.

Additionally, the results presented here were from simulations that were all based on the same data set. We did not investigate how these seven frameworks may fare for different real data sets. Nevertheless, the data set chosen for our empirical study contained mixed variable types, randomization, longitudinal aspects and missing information (including censoring), thereby capturing many realistic and complex features that would be seen in a typical clinical or epidemiological study. It would be interesting to test whether the R-Vine Copula Sequential framework still performs the best for RCT data with different structures, such as time-varying treatments, a more complex outcome (such as a time-to-event outcome), a smaller sample size, or a collection of variables with more complex distributions.

Furthermore, in our sequential frameworks that utilized regression models to produce post-baseline variables, observations with missing values were omitted when variables included in the regression execution model had missingness. In this case, missing values were not generated in the synthetic sample when the original data has missing values. This led to a loss of information and perhaps hindered the ability of the regression models to accurately model the true data distribution. This is because in the real data, observations might not be be missing completely at random and thus the generated distribution mimics that of the observed distribution, but not the true data distribution that would be observed in the absence of missingness (or in the case where observations were missing completely at random). Though not implemented in this paper, we have considered how our sequential framework can still take into account missing values, and even generate missing values. Take, for instance, the scenario in the ACTG175 data where CD4 count at week 96 had non-negligible missingness. A new random variable could be defined, say $R$, where $R = 1$ if the observation for CD4 week 96 was observed, and $R = 0$ if the observation was missing. Then, after generating CD4 count at week 20 but before generating CD4 count at week 96, another execution model could be fit to generate synthetic $R$, i.e., model $\text{Pr}(R=1|\text{Baseline, Treatment, CD4 Week 20})$. This could easily be modeled via logistic regression. Then, to generate CD4 count at week 96, a \textit{weighted} regression execution model could be fit with baseline covariates, treatment assignment, and CD4 count at week 20 included as predictors and weights equal to the inverse of the probability of being observed (estimated by the logistic regression execution model for $R$). In other words, observations that are more likely to be missing are up-weighted. To generate missingness in CD4 count at week 96, we could simply omit observations in the synthetic data for participants with synthetic $R=0$. The same process could be repeated to estimate the weights to be used in a weighted regression execution model when generating the outcome. While we have here outlined a way to incorporate a model for missingness and generate data with missing values, it would be of interest to further explore this approach in the sequential data generation framework for tabular RCT data. Model selection when developing a sequential data generation framework is another intriguing avenue of research to consider.

On the basis of the findings presented here, we recommend utilizing a vine copula approach, in particular R-vine copulas, to generate more complex tables of data (e.g., a set of baseline variables), whereas simpler regression models seem to be sufficient in generating single variables at a time to generate more realistic and meaningful synthetic data sets. Further, we recommend the use of simple statistical comparisons of univariate and bivariate distributions and careful handling of any bounded covariates to ensure faithfulness of the synthetic data to the original data on which they are based. These powerful yet simple approaches to generating realistic synthetic data can be harnessed to study design operating characteristics for new RCTs, to perform more realistic simulations when assessing finite sample performance of new methodology, or to realize "nearly" real data alongside new code without risking any breaches of data confidentiality.

\section*{Author contributions}

NZP contributed to the experimental design, coding of simulations, and writing and editing of the original manuscript. EEMM and NS contributed to the experimental design and editing of the original manuscript.

\section*{Acknowledgments}
EEMM is a Canada Research Chair (Tier 1) in Statistical Methods for Precision Medicine and acknowledges the support of a chercheur de m\'erite career award from the Fonds de Recherche du Qu\'ebec, Sant\'e.
This work is supported by a Discovery Grant from the Natural Sciences and Engineering Research Council of Canada.

\section*{Financial disclosure}

None reported.

\section*{Conflict of interest}

The authors declare no potential conflict of interests.

\section*{Supporting information}

Additional information consisting of a table of the ACTG 175 trial variables used in the analysis as well as further synthetic and real data distribution plots and ML efficacy metric comparisons can be found in Appendix at the end of this article.

\bibliographystyle{apalike}  
\bibliography{references}  

\begin{thebibliography}{}

\bibitem[sdm, 2024]{sdmetrics}
 (2024).
\newblock {\em Synthetic data metrics}.
\newblock DataCebo, Inc.
\newblock Version 0.15.0.

\bibitem[Aas et~al., 2009]{Aas2009}
Aas, K., Czado, C., Frigessi, A., and Bakken, H. (2009).
\newblock Pair-copula constructions of multiple dependence.
\newblock {\em {Insur. Math. Econ}}, 44(2):182--198.

\bibitem[Ainslie et~al., 2018]{Ainslie2018}
Ainslie, K. E.~C., Haber, M.~J., Malosh, R.~E., Petrie, J.~G., and Monto, A.~S. (2018).
\newblock Maximum likelihood estimation of influenza vaccine effectiveness against transmission from the household and from the community.
\newblock {\em Stat. Med}, 37(6):970--983.

\bibitem[Akbari and Liang, 2018]{Akbari2018}
Akbari, M. and Liang, J. (April 15--20, 2018).
\newblock Semi--recurrent {CNN}--based {VAE--GAN} for sequential data generation.
\newblock 2018 IEEE International Conference on Acoustics, Speech and Signal Processing (ICASSP), pages 2321--2325, Calgary, {Canada}. IEEE.

\bibitem[Askin et~al., 2023]{Askin2023}
Askin, S., Burkhalter, D., Calado, G., and El~Dakrouni, S. (2023).
\newblock Artificial intelligence applied to clinical trials: {Opportunities} and challenges.
\newblock {\em {Health Technol}}, 13:203--213.

\bibitem[Assefa et~al., 2020]{Assefa2020}
Assefa, S.~A., Dervovic, D., Mahfouz, M., Tillman, R.~E., Reddy, P., and Veloso, M. (October 15--16, 2020).
\newblock Generating synthetic data in finance: {O}pportunities, challenges and pitfalls.
\newblock ICAIF '20: Proceedings of the First ACM International Conference on AI in Finance, pages 1--8, New York, {United States}. Association for Computing Machinery.

\bibitem[Bedford and Cooke, 1996]{BedfordCooke2002}
Bedford, T. and Cooke, R.~M. (1996).
\newblock Vines - {A} new graphical model for dependent random variables.
\newblock {\em {Ann. Stat}}, 30(4):1031--1068.

\bibitem[Boden-Albala, 2022]{BodenAlbala2022}
Boden-Albala, B. (2022).
\newblock Confronting legacies of underrepresentation in clinical trials: {The} case for greater diversity in research.
\newblock {\em {Neuron}}, 110(5):746--748.

\bibitem[Boren et~al., 2014]{Boren2014}
Boren, D., Sullivan, P.~S., Beyrer, C., Baral, S.~D., Bekker, L.~G., and Brookmeyer, R. (2014).
\newblock Stochastic variation in network epidemic models: {I}mplications for the design of community level {HIV} prevention trials.
\newblock {\em Stat. Med}, 33(22):3894--3904.

\bibitem[Bunkhumpornpat et~al., 2009]{Bunkhumpornpat2009}
Bunkhumpornpat, C., Sinapiromsaran, K., and Lursinsap, C. (April 27--30, 2009).
\newblock {Safe-Level-SMOTE}: {S}afe-level-synthetic minority over-sampling technique for handling the class imbalanced problem.
\newblock Advances in Knowledge Discovery and Data Mining: 13th Pacific-Asia Conference, pages 475--482, Bangkok, {Thailand}. Springer.

\bibitem[Chawla et~al., 2002]{Chawla2002}
Chawla, N.~V., Bowyer, K.~W., Hall, L.~O., and Kegelmeyer, W.~P. (2002).
\newblock Smote: Synthetic minority over-sampling technique.
\newblock {\em {J. Artif. Intell. Res}}, 16:321--357.

\bibitem[Chen et~al., 2024]{Chen2024}
Chen, Q., Ye, A., Zhang, Y., Chen, J., and Huang, C. (2024).
\newblock An intra-class distribution-focused generative adversarial network approach for imbalanced tabular data learning.
\newblock {\em {IJMLC}}, 15(7):2551--2572.

\bibitem[Chen et~al., 2021]{Chen2021}
Chen, Z., Zhang, H., Guo, Y., George, T.~J., Prosperi, M., Hogan, W.~R., He, Z., Shenkman, E.~A., Wang, F., and Bian, J. (2021).
\newblock Exploring the feasibility of using real-world data from a large clinical data research network to simulate clinical trials of {A}lzheimer’s disease.
\newblock {\em {npj Digit. Med}}, 4(1):1--9.

\bibitem[Choi et~al., 2017]{Choi2017}
Choi, E., Biswal, S., Malin, B., Duke, J., Stewart, W.~F., and Sun, J. (August 18--19, 2017).
\newblock Generating multi-label discrete patient records using generative adversarial networks.
\newblock Proceedings of the 2nd Machine Learning for Healthcare Conference, pages 286--305, Boston, {United States}. PMLR.

\bibitem[Collins et~al., 2007]{Collins2007}
Collins, L.~M., Murphy, S.~A., and Strecher, V. (2007).
\newblock The multiphase optimization strategy (most) and the sequential multiple assignment randomized trial (smart): New methods for more potent ehealth interventions.
\newblock {\em Am J Prev Med}, 32(5):112--118.

\bibitem[Crowther and Lambert, 2013]{Crowther2013}
Crowther, M.~J. and Lambert, P.~C. (2013).
\newblock Simulating biologically plausible complex survival data.
\newblock {\em {Stat. Med}}, 32(23):4118--4134.

\bibitem[Dahmen and Cook, 2019]{Dahmen2019}
Dahmen, J. and Cook, D. (2019).
\newblock {SynSys: A} synthetic data generation system for healthcare applications.
\newblock {\em {Sensors}}, 19(5):1--11.

\bibitem[Demeulemeester et~al., 2023]{Demeulemeester2023}
Demeulemeester, R., Savy, N., Grosclaude, P., Costa, N., and Saint-Pierre, P. (2023).
\newblock Agent based modeling in health care economics: {E}xamples in the field of thyroid cancer.
\newblock {\em {Int. J. Biostat}}, 19(2):351--368.

\bibitem[Denton et~al., 2015]{Denton2015}
Denton, E.~L., Chintala, S., Szlam, A., and Fergus, R. (December 7--10, 2015).
\newblock Deep generative image models using a {Laplacian} pyramid of adversarial networks.
\newblock Advances in Neural Information Processing Systems, pages 1--9, Montreal, {Canada}. Curran Associates, Inc.

\bibitem[Douzas et~al., 2018]{Douzas2018}
Douzas, G., Bacao, F., and Last, F. (2018).
\newblock Improving imbalanced learning through a heuristic oversampling method based on k-means and {SMOTE}.
\newblock {\em {Inf. Sci}}, 465:1--20.

\bibitem[Figueira and Vaz, 2022]{Figueira2022}
Figueira, A. and Vaz, B. (2022).
\newblock Survey on synthetic data generation, evaluation methods and {GANs}.
\newblock {\em {Mathematics}}, 10(15):1--41.

\bibitem[Franklin et~al., 2014]{Franklin2014}
Franklin, J.~M., Schneeweiss, S., Polinski, J.~M., and Rassen, J.~A. (2014).
\newblock Plasmode simulation for the evaluation of pharmacoepidemiologic methods in complex healthcare databases.
\newblock {\em {Comput. Stat. Data Anal}}, 72:219--226.

\bibitem[Genest et~al., 2024]{Genest2024}
Genest, C., Okhrin, O., and Bodnar, T. (2024).
\newblock Copula modeling from {Abe Sklar} to the present day.
\newblock {\em {JMVA}}, 201:1--9.

\bibitem[Goodfellow et~al., 2020]{Goodfellow2014}
Goodfellow, I., Pouget-Abadie, J., Mirza, M., Xu, B., Warde-Farley, D., Ozair, S., Courville, A., and Bengio, Y. (2020).
\newblock Generative adversarial networks.
\newblock {\em {Commun. ACM}}, 63(11):139--144.

\bibitem[Hammer et~al., 1996]{Hammer1996}
Hammer, S.~M., Katzenstein, D.~A., Hughes, M.~D., Gundacker, H., Schooley, R.~T., Haubrich, R.~H., Henry, K.~W., Lederman, M.~M., Phair, J.~P., Niu, M., Hirsch, M.~S., and Merigan, T.~C. (1996).
\newblock A trial comparing nucleoside monotherapy with combination therapy in {HIV-infected} adults with {CD4} cell counts from 200 to 500 per cubic millimeter.
\newblock {\em {NEJM}}, 335(15):1081--1090.

\bibitem[Han et~al., 2005]{Han2005}
Han, H., Wang, W.~Y., and Mao, B.~H. (August 23--26, 2005).
\newblock {Borderline-SMOTE}: {A} new over-sampling method in imbalanced data sets learning.
\newblock International Conference on Intelligent Computing, pages 878--887, Hefei, {China}. Springer.

\bibitem[Hastie et~al., 2001]{Hastie2001}
Hastie, T., Tibshirani, R., Sherlock, G., Eisen, M., Brown, P., and Botstein, D. (2001).
\newblock Imputing missing data for gene expression arrays.
\newblock {\em Technical report, Stanford Statistics Department}, 17(6):520--525.

\bibitem[He et~al., 2008]{He2008}
He, H., Bai, Y., Garcia, E.~A., and Li, S. (June 1--8, 2008).
\newblock Adasyn: Adaptive synthetic sampling approach for imbalanced learning.
\newblock 2008 IEEE International Joint Conference on Neural Networks (IEEE World Congress on Computational Intelligence), pages 1322--1328, Hong Kong, {China}. IEEE.

\bibitem[Heiat et~al., 2002]{Heiat2002}
Heiat, A., Gross, C.~P., and Krumholz, H.~M. (2002).
\newblock Representation of the elderly, women, and minorities in heart failure clinical trials.
\newblock {\em {AMA Arch. Intern. Med}}, 162(15):1682--1688.

\bibitem[Hernandez et~al., 2022]{Hernandez2022}
Hernandez, M., Epelde, G., Alberdi, A., Cilla, R., and Rankin, D. (2022).
\newblock Synthetic data generation for tabular health records: {A} systematic review.
\newblock {\em {Neurocomputing}}, 493:28--45.

\bibitem[Hsu et~al., 2020]{Hsu2020}
Hsu, A., Khoo, W., Goyal, N., and Wainstein, M. (2020).
\newblock Next-generation digital ecosystem for climate data mining and knowledge discovery: {A} review of digital data collection technologies.
\newblock {\em {Front. Big Data}}, 3(29):1--19.

\bibitem[Jeni et~al., 2013]{Jeni2013}
Jeni, L.~A., Cohn, J.~F., and {De La Torre}, F. (September 02--05, 2013).
\newblock Facing imbalanced data--recommendations for the use of performance metrics.
\newblock 2013 Humaine Association Conference on Affective Computing and Intelligent Interaction, pages 245--251, Geneva, {Switzerland}. IEEE.

\bibitem[Joe, 1996]{Joe1996}
Joe, H. (1996).
\newblock Families of m-variate distributions with given margins and m (m-1)/2 bivariate dependence parameters.
\newblock {\em Lecture notes-monograph series}, 28:120--141.

\bibitem[Kingma and Welling, 2013]{Kingma2013}
Kingma, D.~P. and Welling, M. (May 2--4, 2013).
\newblock Auto-encoding variational {B}ayes.
\newblock International Conference on Learning Representations 2013, pages 1--14, Scottsdale, {United States}. CoRR.

\bibitem[Kiran and Kumar, 2024]{Kiran2024}
Kiran, A. and Kumar, S.~S. (2024).
\newblock A methodology and an empirical analysis to determine the most suitable synthetic data generator.
\newblock {\em {IEEE Access}}, 12:12209--12228.

\bibitem[Kiran and Kumar, 2023]{Kiran2023}
Kiran, A. and Kumar, S.~S. (March 3--5, 2023).
\newblock A comparative analysis of gan and vae based synthetic data generators for high dimensional, imbalanced tabular data.
\newblock 2023 2nd International Conference for Innovation in Technology (INOCON), pages 1--6, Bangalore, {India}. IEEE.

\bibitem[Maglio and Mabry, 2011]{Maglio2011}
Maglio, P.~P. and Mabry, P.~L. (2011).
\newblock Agent-based models and systems science approaches to public health.
\newblock {\em Am. J. Prev. Med}, 40(3):392--394.

\bibitem[Mirza and Osindero, 2014]{Mirza2014}
Mirza, M. and Osindero, S. (2014).
\newblock Conditional generative adversarial nets.
\newblock Accessed September 07, 2024. \url{http://arxiv.org/abs/1411.1784}, {arXiv}, 1411.1784.

\bibitem[Nagler and Vatter, 2023]{rvinecopulib}
Nagler, T. and Vatter, T. (2023).
\newblock {\em rvinecopulib: High performance algorithms for vine copula modeling}.
\newblock Version 0.6.3.1.1.

\bibitem[O'Keefe and Rubin, 2015]{OKeefe2015}
O'Keefe, C.~M. and Rubin, D.~B. (2015).
\newblock Individual privacy versus public good: {P}rotecting confidentiality in health research.
\newblock {\em Stat. Med}, 34(23):3081--3103.

\bibitem[Pappalardo et~al., 2019]{Pappalardo2019}
Pappalardo, F., Russo, G., Tshinanu, F.~M., and Viceconti, M. (2019).
\newblock In silico clinical trials: {Concepts} and early adoptions.
\newblock {\em {Brief Bioinform}}, 20(5):1699--1708.

\bibitem[Patki et~al., 2016]{Patki2016}
Patki, N., Wedge, R., and Veeramachaneni, K. (October 17--19, 2016).
\newblock The synthetic data vault.
\newblock 2016 {IEEE} International Conference on Data Science and Advanced Analytics {(DSAA)}, pages 399--410, Montreal, {Canada}. IEEE.

\bibitem[Pezoulas et~al., 2024]{Pezoulas2024}
Pezoulas, V.~C., Zaridis, D.~I., Mylona, E., Androutsos, C., Apostolidis, K., Tachos, N.~S., and Fotiadis, D.~I. (2024).
\newblock Synthetic data generation methods in healthcare: {A} review on open-source tools and methods.
\newblock {\em {CSBJ}}, 23:2892--2910.

\bibitem[Radford et~al., 2015]{Radford2015}
Radford, A., Metz, L., and Chintala, S. (May 7--9, 2015).
\newblock Unsupervised representation learning with deep convolutional generative adversarial networks.
\newblock International Conference on Learning Representations 2015, pages 1--16, San Diego, {United States}. CoRR.

\bibitem[Rajabi and Garibay, 2022]{Rajabi2022}
Rajabi, A. and Garibay, O.~O. (2022).
\newblock {TabFairGAN: Fair} tabular data generation with generative adversarial networks.
\newblock {\em {Mach. Learn. Knowl. Extr}}, 4(2):488--501.

\bibitem[Saint-Pierre and Savy, 2023]{SaintPierre2023}
Saint-Pierre, P. and Savy, N. (2023).
\newblock Agent-based modeling in medical research, virtual baseline generator and change in patients’ profile issue.
\newblock {\em {Int. J. Biostat}}, 19(2):333--349.

\bibitem[Sarrami-Foroushani et~al., 2021]{SarramiForoushani2021}
Sarrami-Foroushani, A., Lassila, T., MacRaild, M., Asquith, J., Roes, K. C.~B., Byrne, J.~V., and Frangi, A.~F. (2021).
\newblock In-silico trial of intracranial flow diverters replicates and expands insights from conventional clinical trials.
\newblock {\em {Nat. Commun}}, 12(1):1--12.

\bibitem[Schreck et~al., 2024]{Schreck2024}
Schreck, N., Slynko, A., Saadati, M., and Benner, A. (2024).
\newblock Statistical plasmode simulations -- {P}otentials, challenges and recommendations.
\newblock {\em {Stat. Med}}, 43(9):1804--1825.

\bibitem[Sklar, 1959]{Sklar1959}
Sklar, A. (1959).
\newblock Fonctions de répartition à n dimmensions et leurs marges.
\newblock {\em Annales de l’ISUP}, 8(3):229--231.

\bibitem[Souli et~al., 2023]{Souli2023}
Souli, Y., Trudel, X., Diop, A., Brisson, C., and Talbot, D. (2023).
\newblock Longitudinal plasmode algorithms to evaluate statistical methods in realistic scenarios: {An} illustration applied to occupational epidemiology.
\newblock {\em {BMC Med Res Methodol}}, 23(242):1--15.

\bibitem[Sun et~al., 2019]{Sun2019}
Sun, Y., Cuesta-Infante, A., and Veeramachaneni, K. (January 21 -- February 1, 2019).
\newblock Learning vine copula models for synthetic data generation.
\newblock Proceedings of the {AAAI} Conference on Artificial Intelligence, pages 5049--5057, Honolulu, {United States}. AAAI.

\bibitem[Vaughan et~al., 2009]{Vaughan2009}
Vaughan, L.~K., Divers, J., Padilla, M.~A., Redden, D.~T., Tiwari, H.~K., Pomp, D., and Allison, D.~B. (2009).
\newblock The use of plasmodes as a supplement to simulations: {A} simple example evaluating individual admixture estimation methodologies.
\newblock {\em {Comput. Stat. Data Anal}}, 53(5):1755--1766.

\bibitem[Walia et~al., 2020]{Walia2020}
Walia, M., Tierney, B., and McKeever, S. (December 7--8, 2020).
\newblock Synthesising tabular data using {Wasserstein} conditional {GANs} with gradient penalty {(WCGAN-GP)}.
\newblock 28th Irish Conference on Artificial Intelligence and Cognitive Science, pages 1--12, Dublin, {Ireland}. AICS.

\bibitem[Wang and Pai, 2023]{Wang2023}
Wang, W. and Pai, T.~W. (2023).
\newblock Enhancing small tabular clinical trial dataset through hybrid data augmentation: {Combining SMOTE and WCGAN-GP}.
\newblock {\em Data}, 8(9):1--20.

\bibitem[Xu et~al., 2019]{Xu2019}
Xu, L., Skoularidou, M., Cuesta-Infante, A., and Veeramachaneni, K. (December 8--14, 2019).
\newblock Modeling tabular data using conditional {GAN}.
\newblock Advances in Neural Information Processing Systems, pages 1--11, Vancouver, {Canada}. NeurIPS.

\bibitem[Yan et~al., 2018]{Yan2018}
Yan, Q.~L., Tang, S.~Y., and Xiao, Y.~N. (2018).
\newblock Impact of individual behaviour change on the spread of emerging infectious diseases.
\newblock {\em Stat. Med}, 37(6):948--969.

\bibitem[Zhou et~al., 2020]{Zhou2020}
Zhou, Y., Dong, F., Liu, Y., Li, Z., Du, J., and Zhang, L. (2020).
\newblock Forecasting emerging technologies using data augmentation and deep learning.
\newblock {\em {Scientometrics}}, 123:1--29.

\bibitem[Zwep et~al., 2024]{Zwep2024}
Zwep, L.~B., Guo, T., Nagler, T., Knibbe, C. A.~J., Meulman, J.~J., and {van Hasselt}, J. G.~C. (2024).
\newblock Virtual patient simulation using copula modeling.
\newblock {\em {Clin. Pharmacol. Ther}}, 115(4):795--804.

\end{thebibliography}






\newpage

\section*{Appendix}

\subsection*{Tables}

\begin{xltabular}{\linewidth}{ l | X }
  \caption{Description of variables used in our investigation of realistic synthetic data generation frameworks.} 
 \label{table: vardescription}\\
 \hline \hline

\textbf{\normalsize Variable Name} & \textbf{\normalsize Definition and Support}  \\
 \hline 
\endfirsthead
 \hline \hline

\textbf{\normalsize Variable Name} & \textbf{\normalsize Definition and support}  \\
 \hline 
\endhead

\textbf{Participant ID} & Unique identifier for each participant. Numeric value. \\ \hline 

\textbf{Age} & Age, in years. Continuous variable, greater than 12. \\ \hline 

\textbf{Weight} & Weight, in kg. Continuous variable, greater than zero. \\ \hline 

\textbf{Sex} & Participant sex. Binary variable (female, male). \\ \hline 

\textbf{Race} & Participant race. Binary variable (white, non-white). \\ \hline 

\textbf{Hemophilia} & Hemophilia status. Binary variable (yes, no). \\ \hline 

\textbf{Homosexuality} & Homosexuality identity. Binary variable (yes, no). \\ \hline 

\textbf{Drug Use} & History of intravenous drug use. Binary variable (yes, no). \\ \hline 

\textbf{Karnofsky Score} & Karnofsky health score. Continuous variable, [0, 100] \\ \hline 

\textbf{Prior Non-Zidovudine ART Usage} & History of non-zidovudine ART usage prior to study. Binary variable (yes, no). \\ \hline 

\textbf{Zidovudine Usage 30 Days Prior} & History of zidovudine usage in the 30 days prior to study treatment initiation. Binary variable (yes, no). \\ \hline

\textbf{Previous Time on ART} & Duration of previously received ART, in days. Continuous variable, greater than or equal to zero. \\ \hline 

\textbf{ART History} & History of ART usage. Categorical variable (naive, 1-52 weeks of prior ART, more than 52 weeks of prior ART). \\ \hline 

\textbf{Symptomatic HIV} & Status of symptomatic HIV. Binary variable (asymptomatic, symptomatic). \\ \hline 

\textbf{CD4 Baseline} & CD4 cell count at baseline. Continuous variable, greater than or equal to zero. \\ \hline 

\textbf{Treatment} & Treatment arms. Categorical variable (zidovudine only, zidovudine and didanosine, zidovudine and zalcitabine, didanosine only). \\ \hline 

\textbf{CD4 Week 20} & CD4 cell count at week 20 visit. Continuous variable, greater than or equal to zero. \\ \hline 

\textbf{CD4 Week 96} & CD4 cell count at week 96 visit. Continuous variable, greater than or equal to zero. \\ \hline 

\textbf{Outcome} & Trial outcome. Binary variable (observed composite event, did not observe composite event). \\ \hline 

\end{xltabular}

\newpage

\subsection*{Figures}

\FloatBarrier
\begin{figure}[h]
    \centering
    \includegraphics[width=\linewidth]{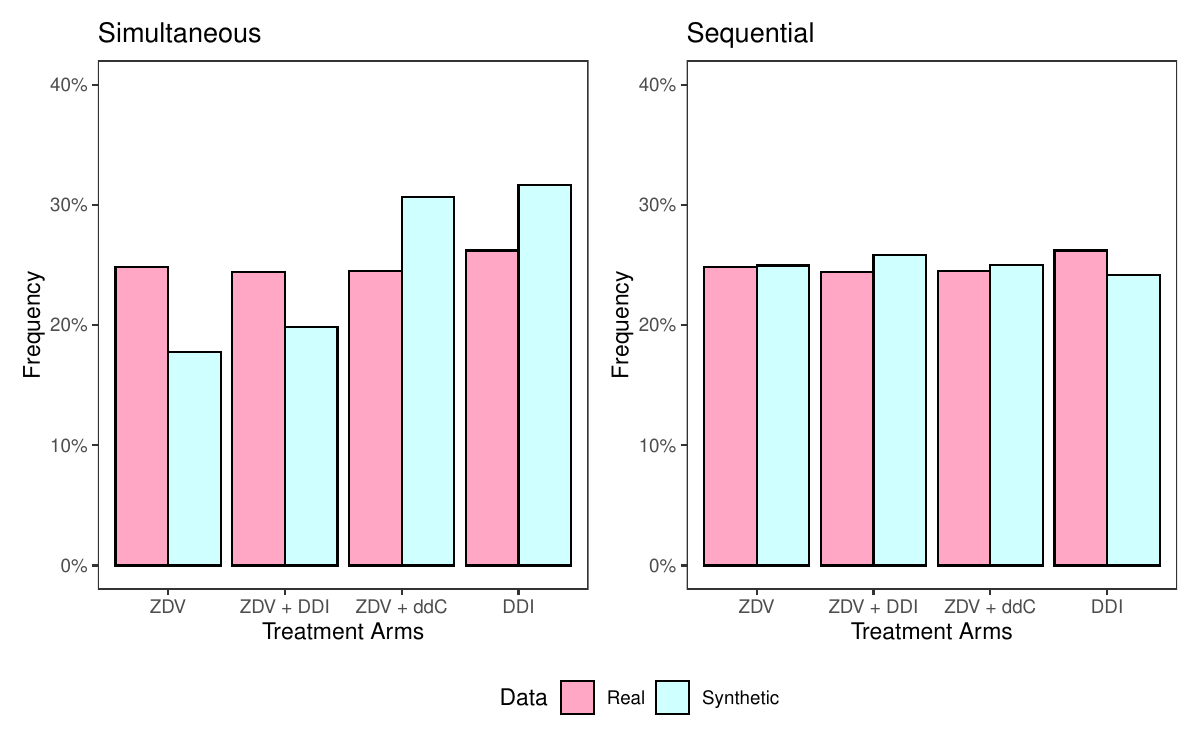}
    \caption{Density plots of treatment assignment. The plot on the left shows the distribution for data generated via the CTGAN Simultaneous framework, and the plot on the right shows the distribution for data generated by the treatment assignment execution model.}
    \label{fig:tx_simult_sequent}
\end{figure}

\begin{figure}
    \centering
    \includegraphics[width=\linewidth]{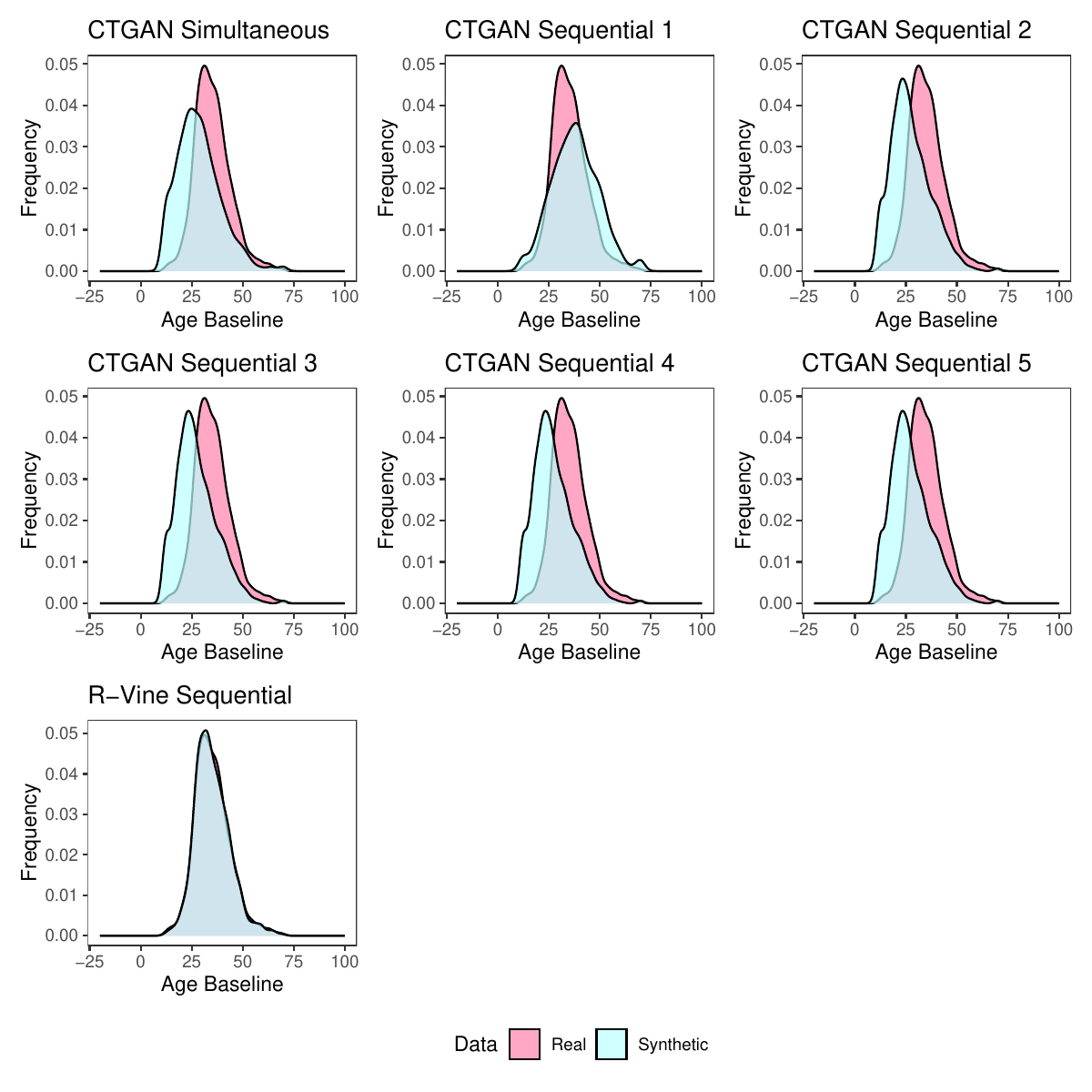}
    \caption{Density plots of age at baseline for a single data generation run as compared to the real data for seven candidate synthetic data generators.}
    \label{fig:7methods_age}
\end{figure}

\begin{figure}
    \centering
    \includegraphics[width=\linewidth]{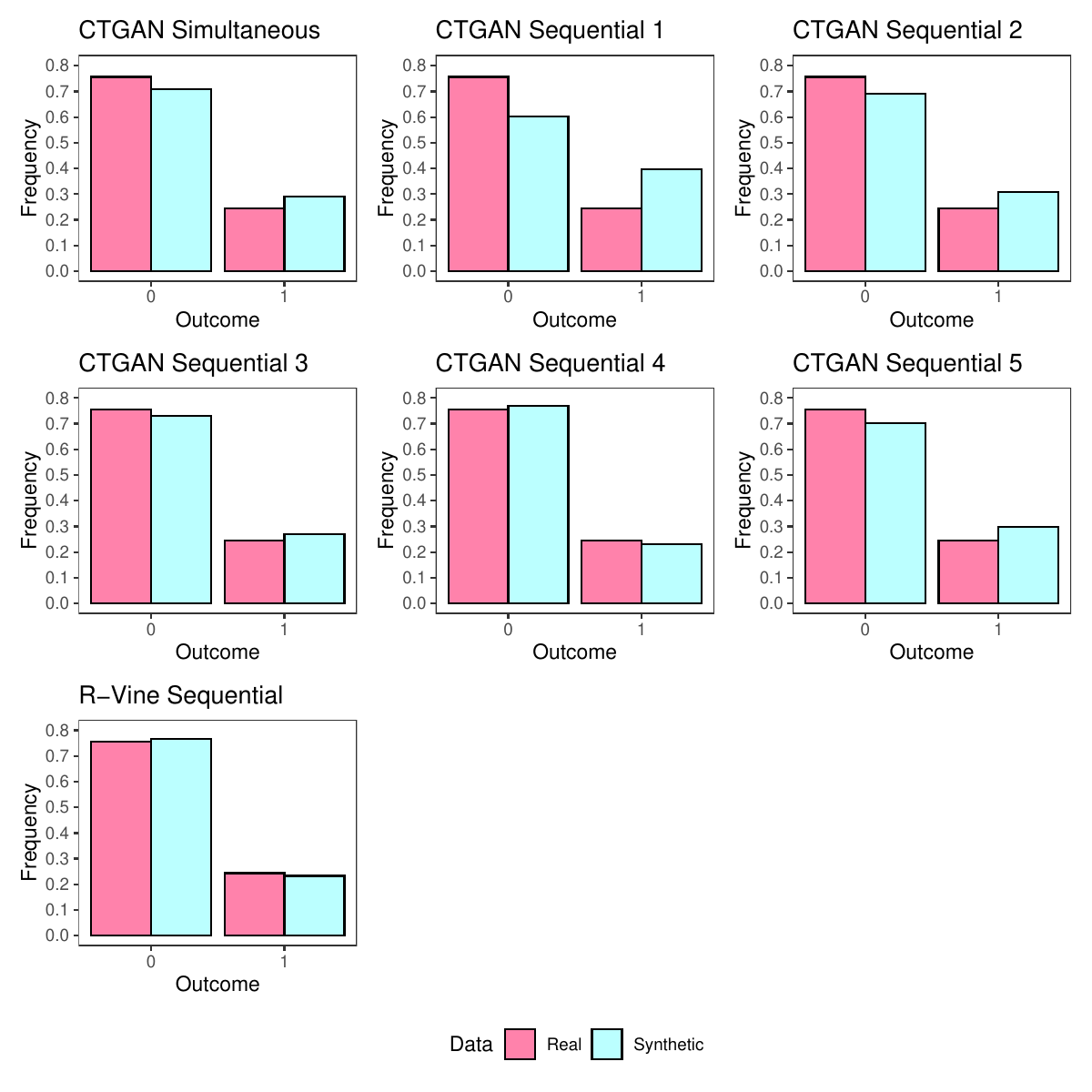}
    \caption{Density plots of the composite event outcome for a single data generation run as compared to the real data for seven candidate synthetic data generators.}
    \label{fig:7methods_outcome}
\end{figure}

\begin{figure}
    \centering
    \includegraphics[width=\linewidth]{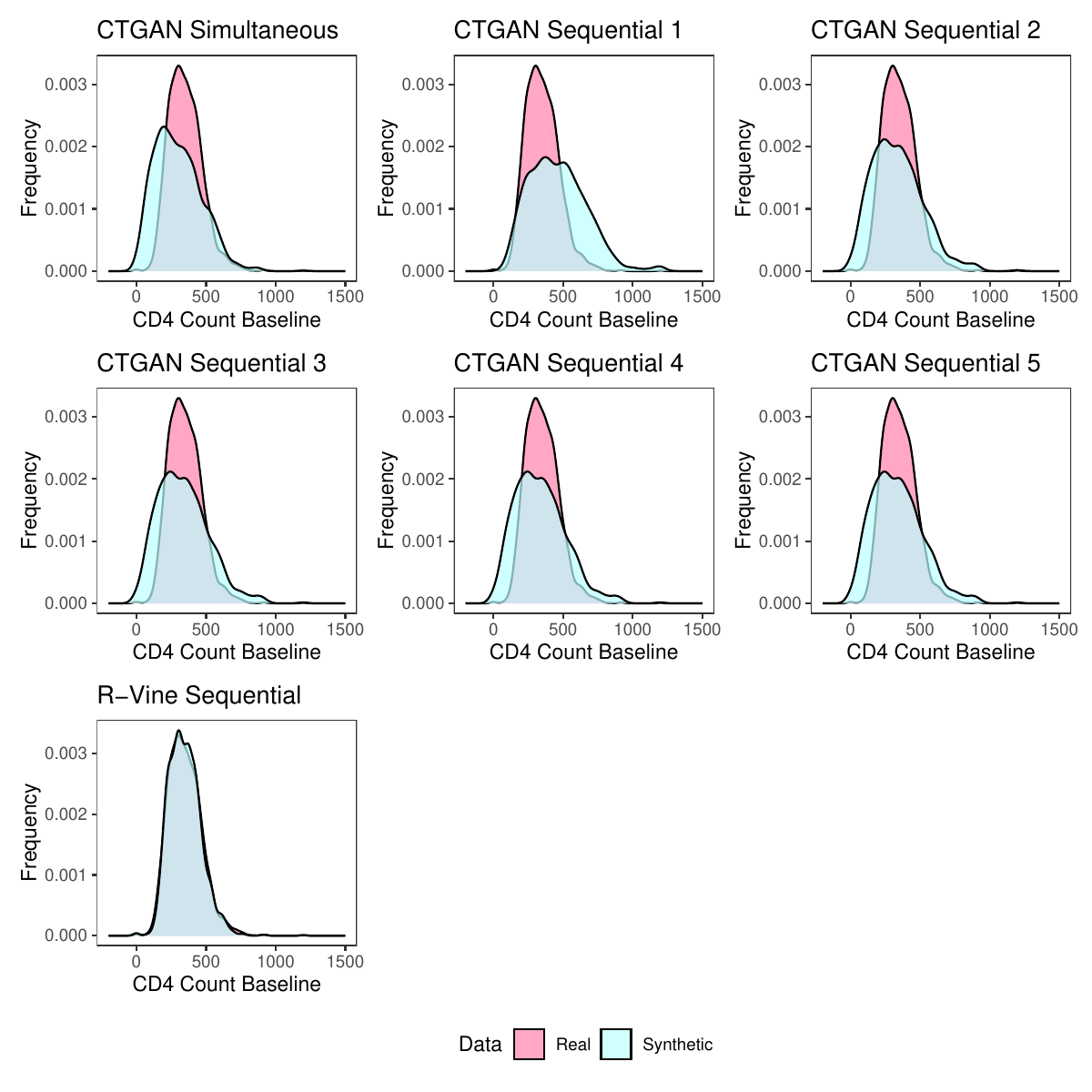}
    \caption{Density plots of CD4 cell count at baseline for a single data generation run as compared to the real data for seven candidate synthetic data generators.}
    \label{fig:7methods_cd40}
\end{figure}

\begin{figure}
    \centering
    \includegraphics[width=\linewidth]{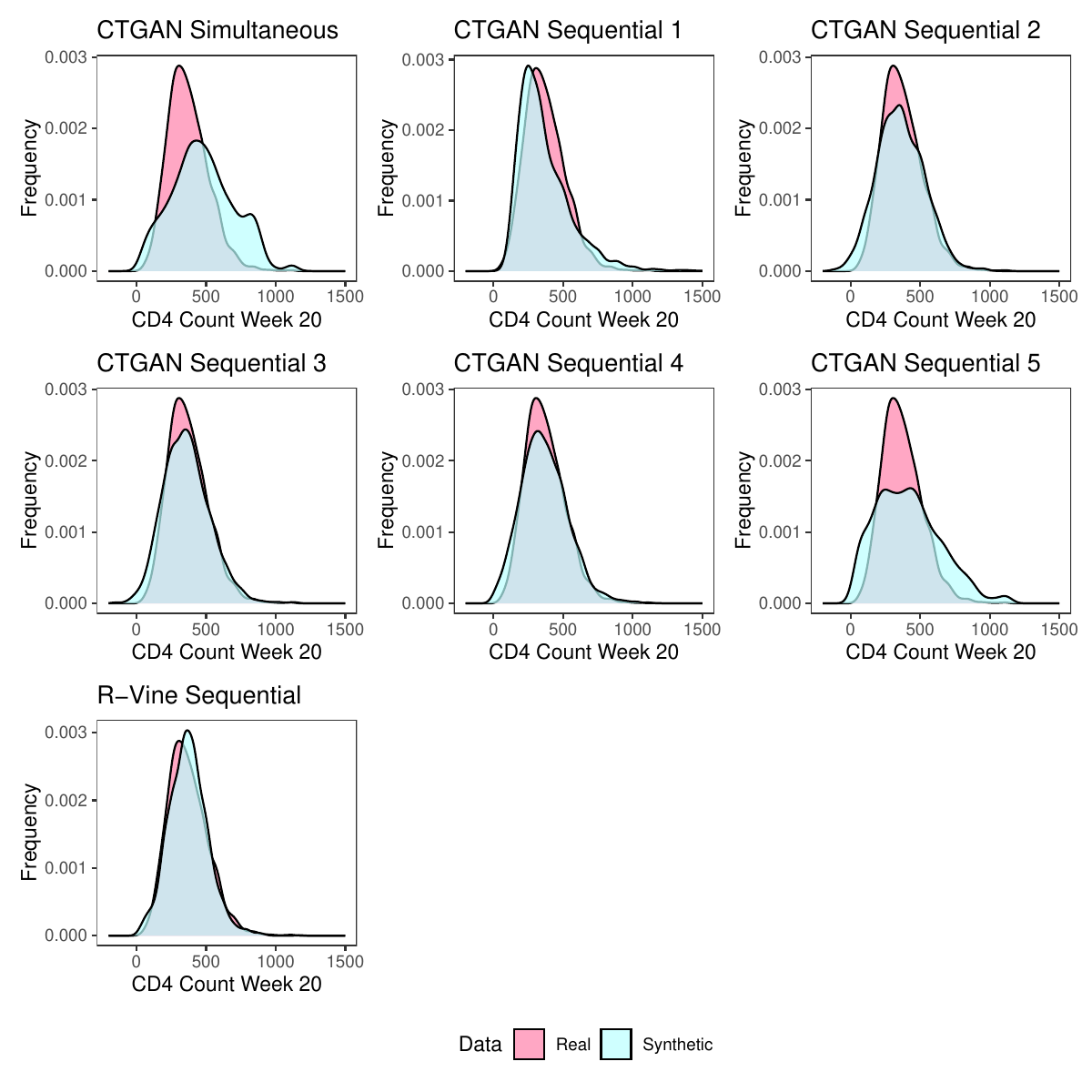}
    \caption{Univariate distribution plots of CD4 cell count at week 20 for a single data generation run as compared to the real data for seven candidate synthetic data generators.}
    \label{fig:7methods_cd420}
\end{figure}

\begin{figure}
    \centering
    \includegraphics[width=\linewidth]{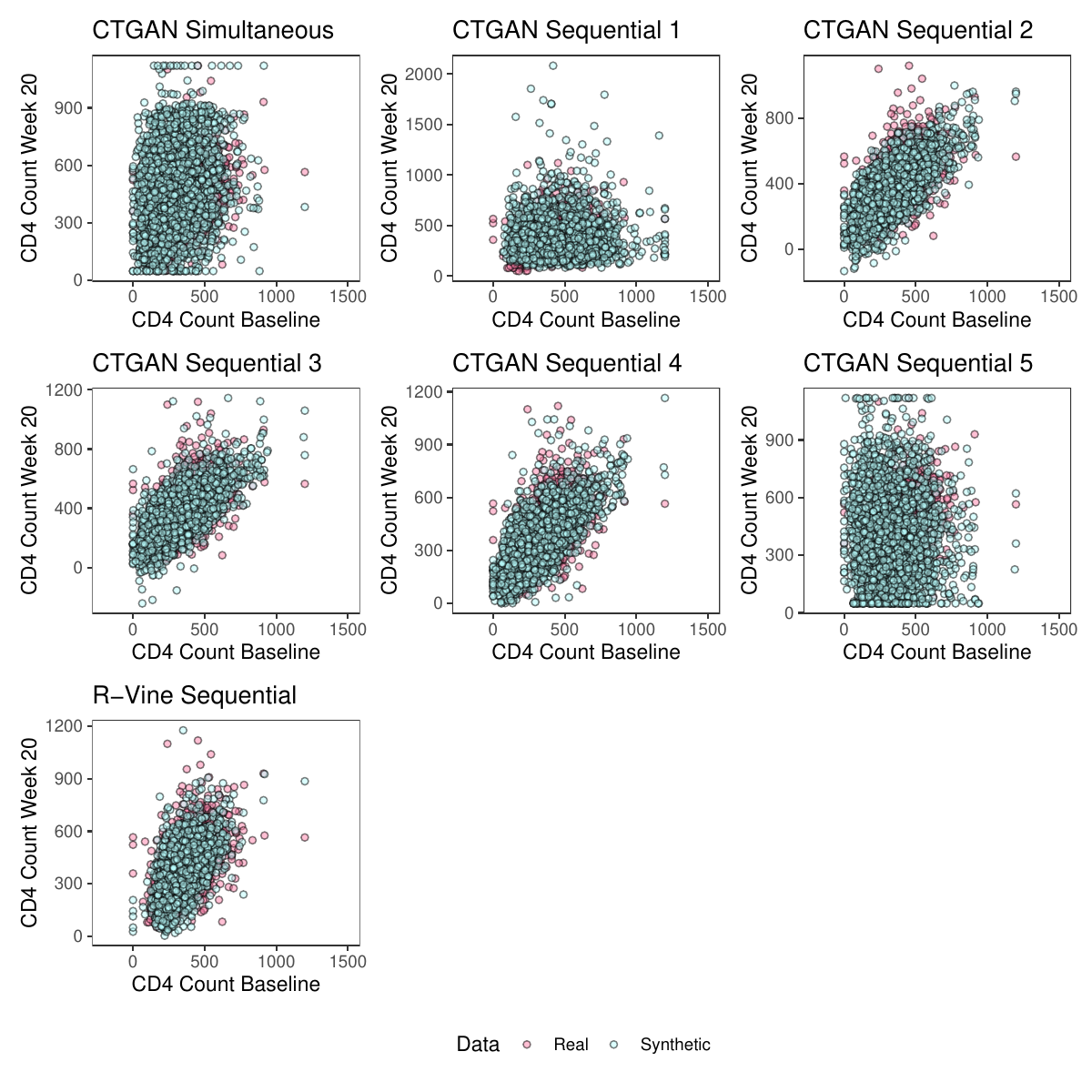}
    \caption{Bivariate distribution plots of CD4 cell count at week 20 against CD4 cell count at baseline for a single data generation run as compared to the real data for seven candidate synthetic data generators.}
    \label{fig:7methods_cd40_cd420}
\end{figure}

\begin{figure}
    \centering
    \includegraphics[width=\linewidth]{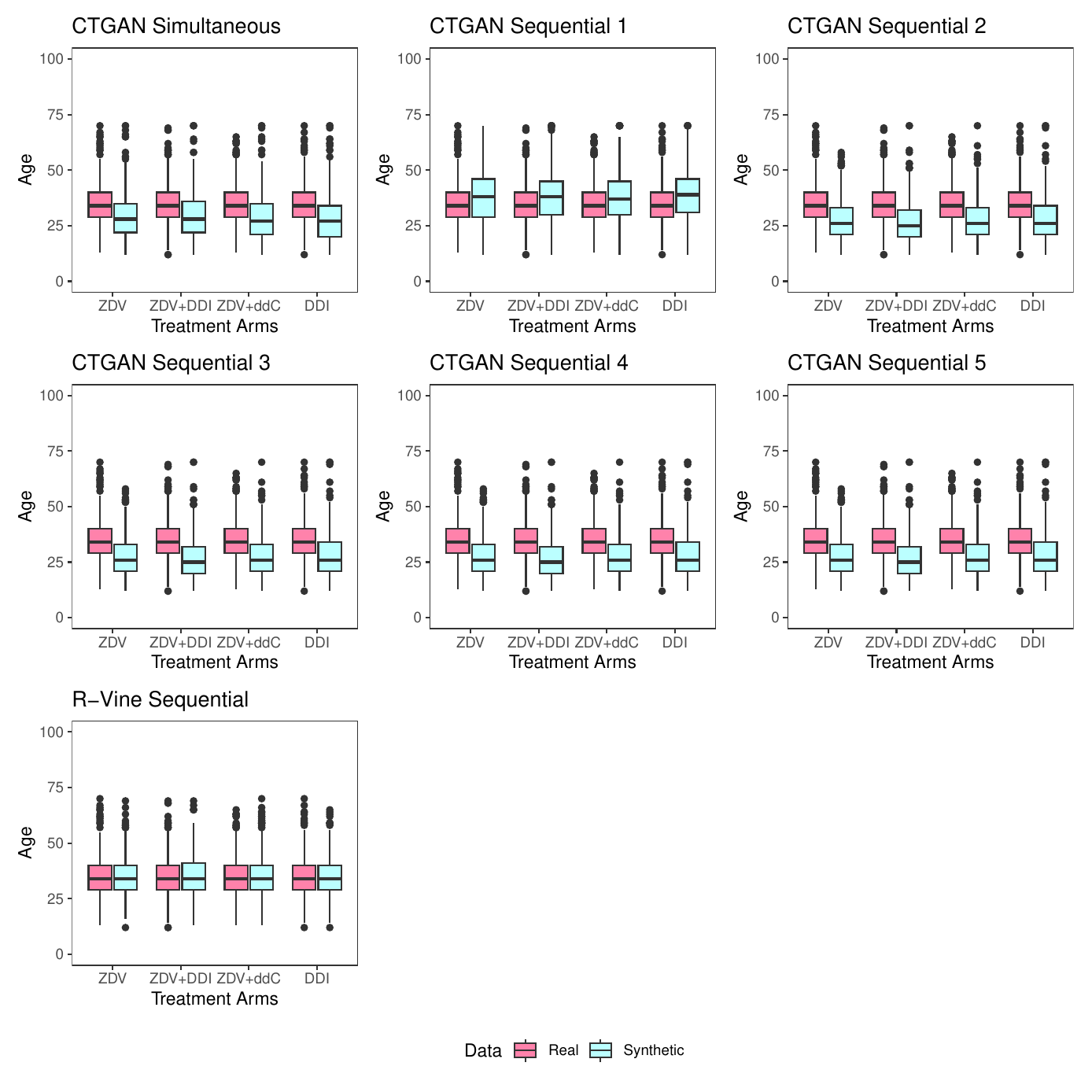}
    \caption{Bivariate distribution plots of age against treatment arm for a single data generation run as compared to the real data for seven candidate synthetic data generators.}
    \label{fig:7methods_tx_age}
\end{figure}

\begin{figure*}[t!]
        \subfloat[Run 1]{%
            \includegraphics[width=.45\linewidth]{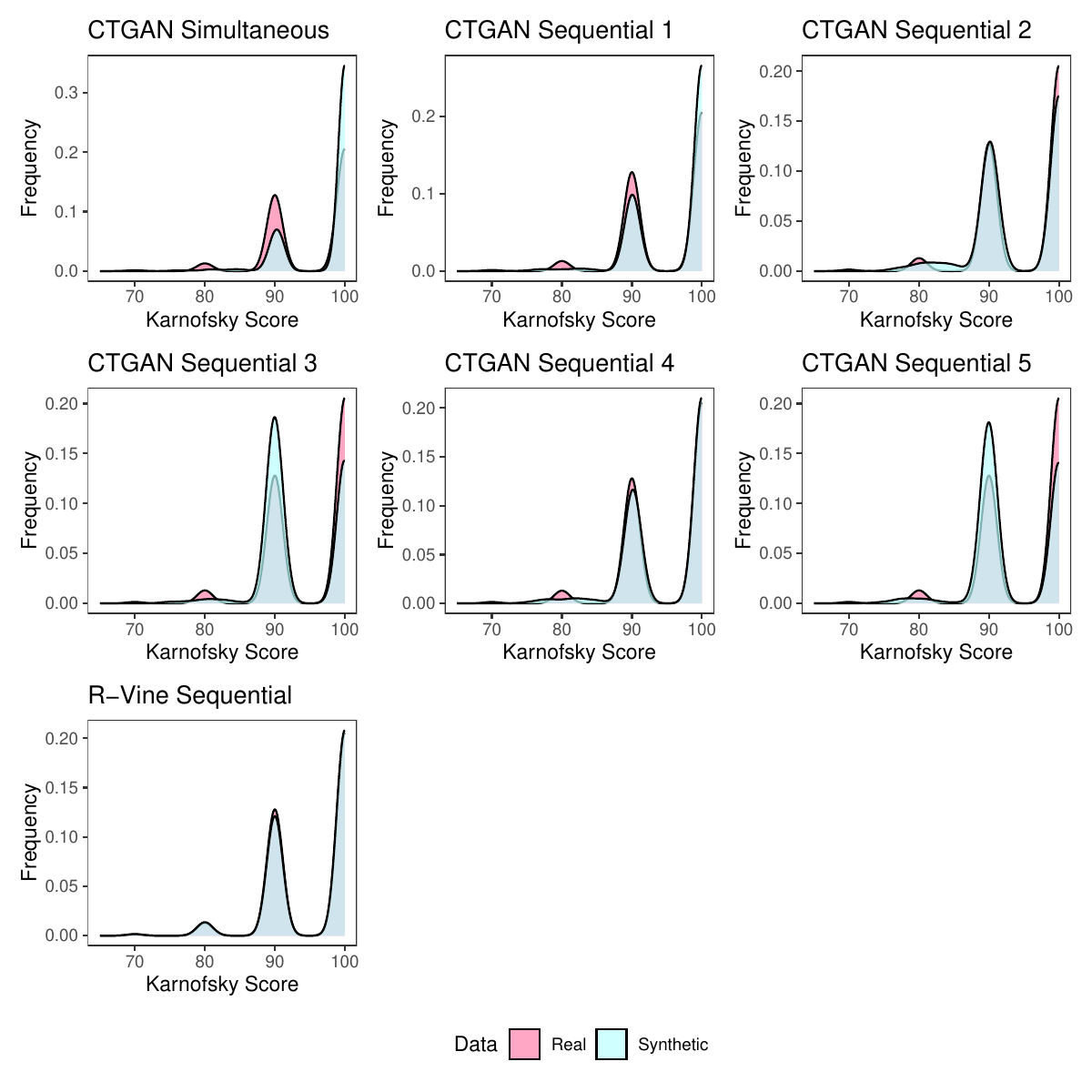}%
            \label{subfig:a}%
        }\hfill
        \subfloat[Run 2]{%
            \includegraphics[width=.45\linewidth]{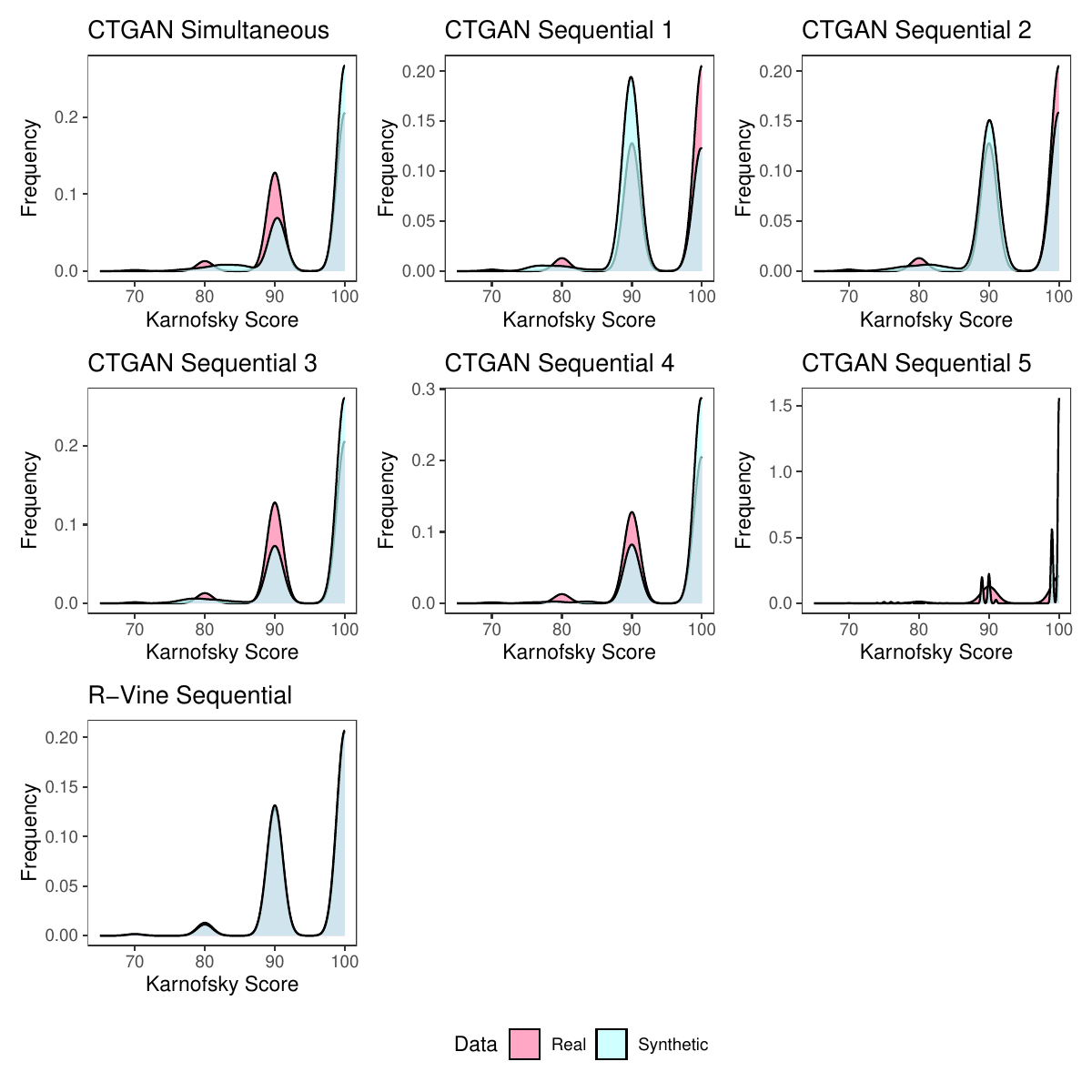}%
            \label{subfig:b}%
        }\\
        \subfloat[Run 3]{%
            \includegraphics[width=.45\linewidth]{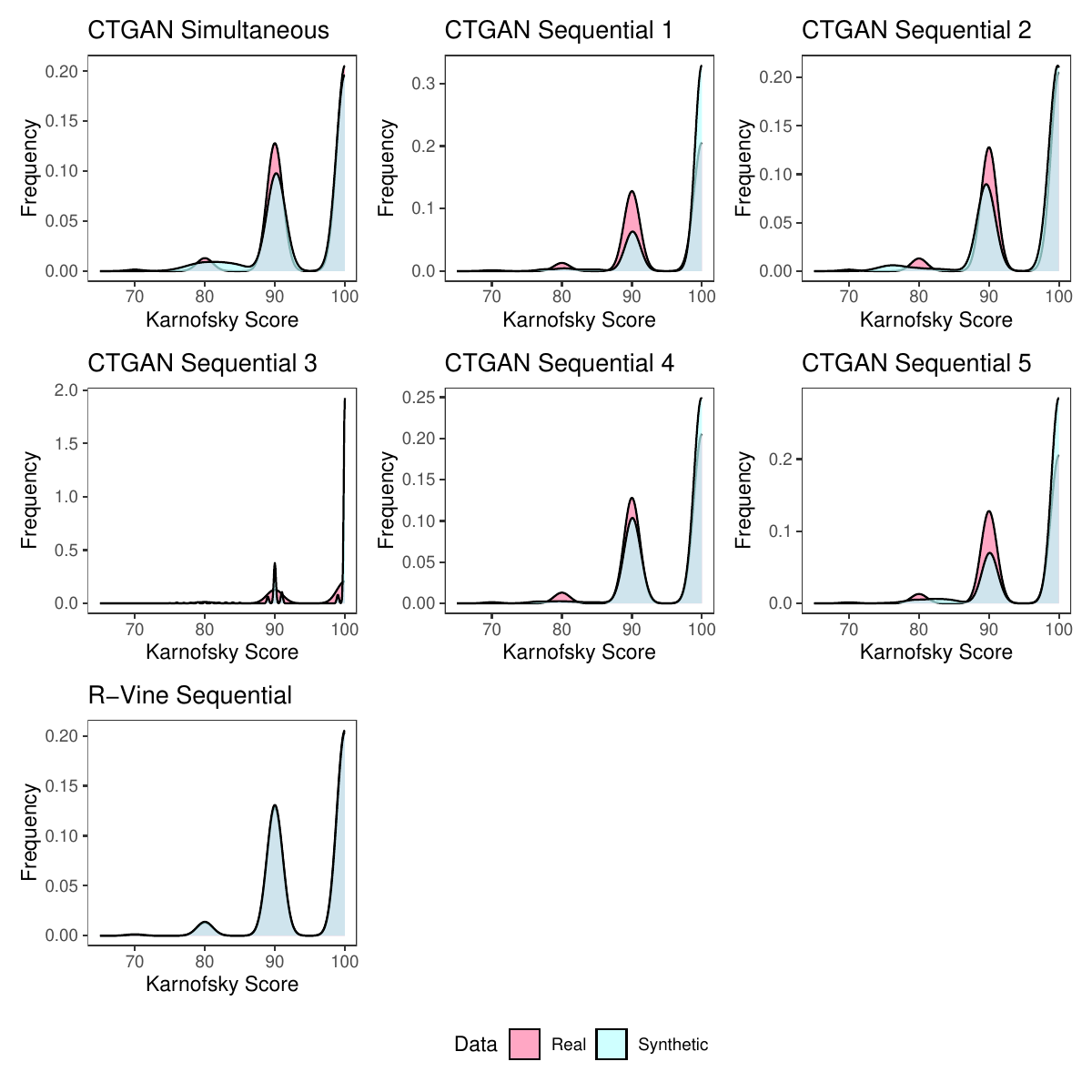}%
            \label{subfig:c}%
        }\hfill
        \subfloat[Run 4]{%
            \includegraphics[width=.45\linewidth]{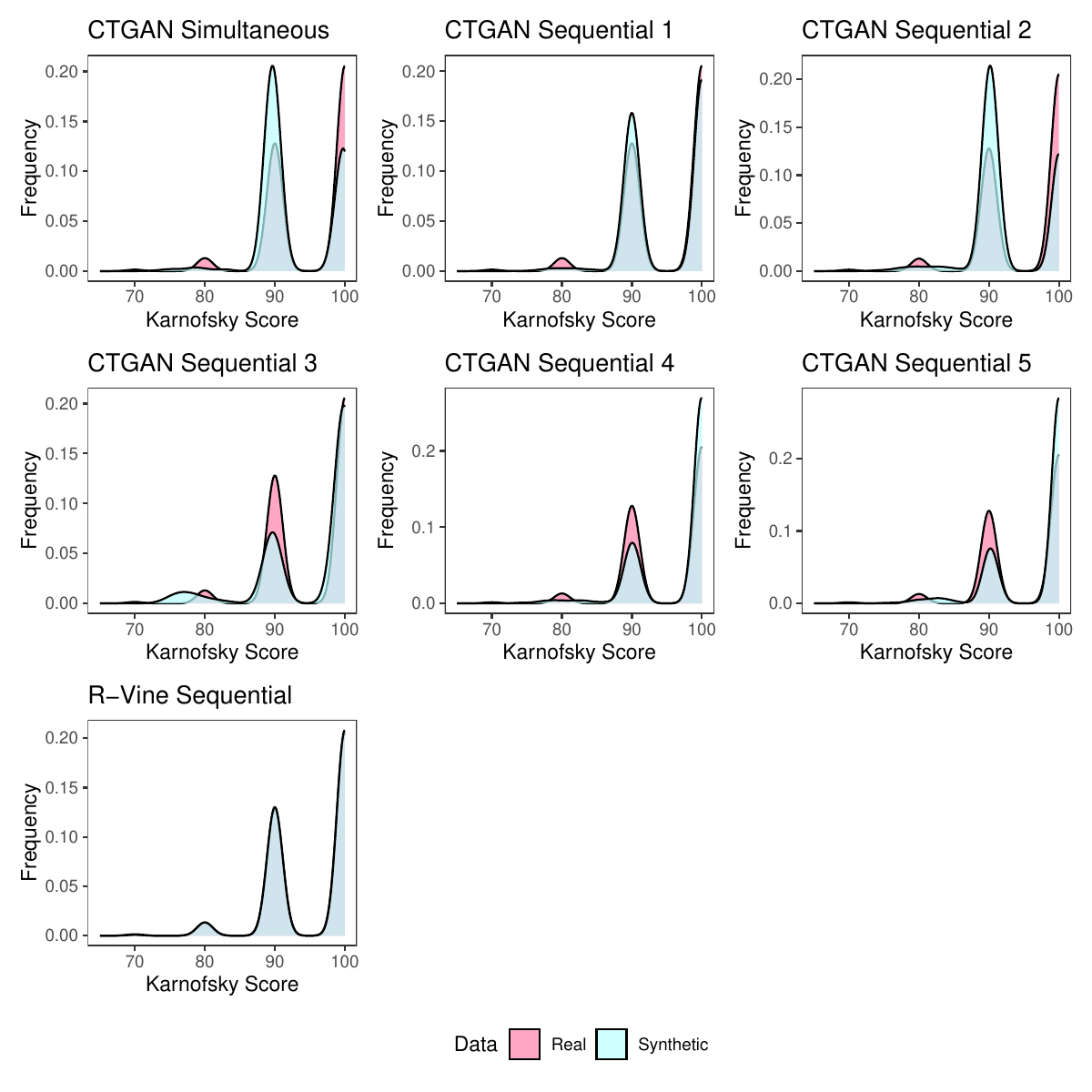}%
            \label{subfig:d}%
        }
        \caption{Density plots (pink for real data, blue for synthetic data) of Karnofsky score across four different data generation runs for seven candidate synthetic data generators.}
        \label{fig:karnof4runs}
\end{figure*}

\begin{figure*}[t!]
        \subfloat[Run 1]{%
            \includegraphics[width=.45\linewidth]{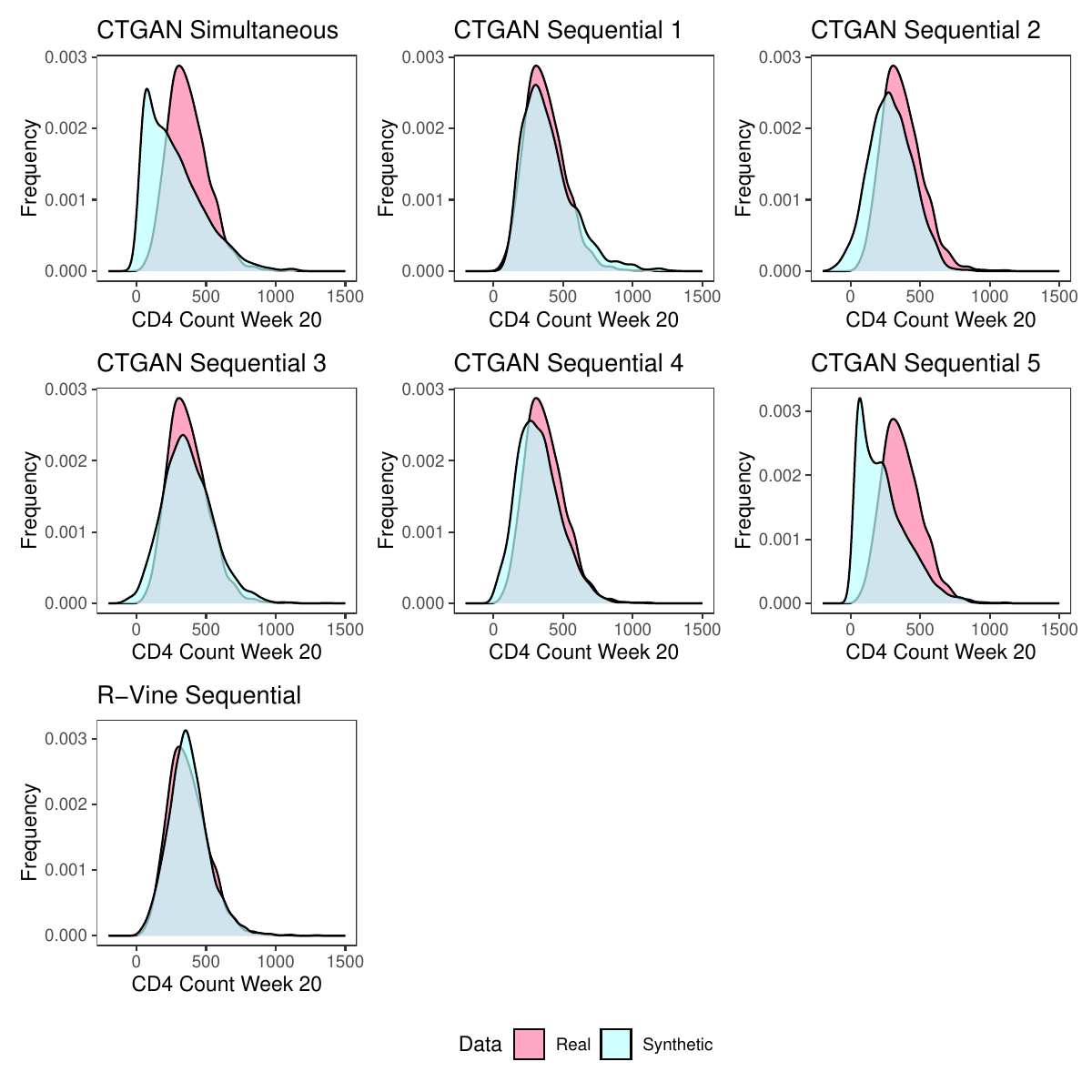}%
            \label{subfig:a}%
        }\hfill
        \subfloat[Run 2]{%
            \includegraphics[width=.45\linewidth]{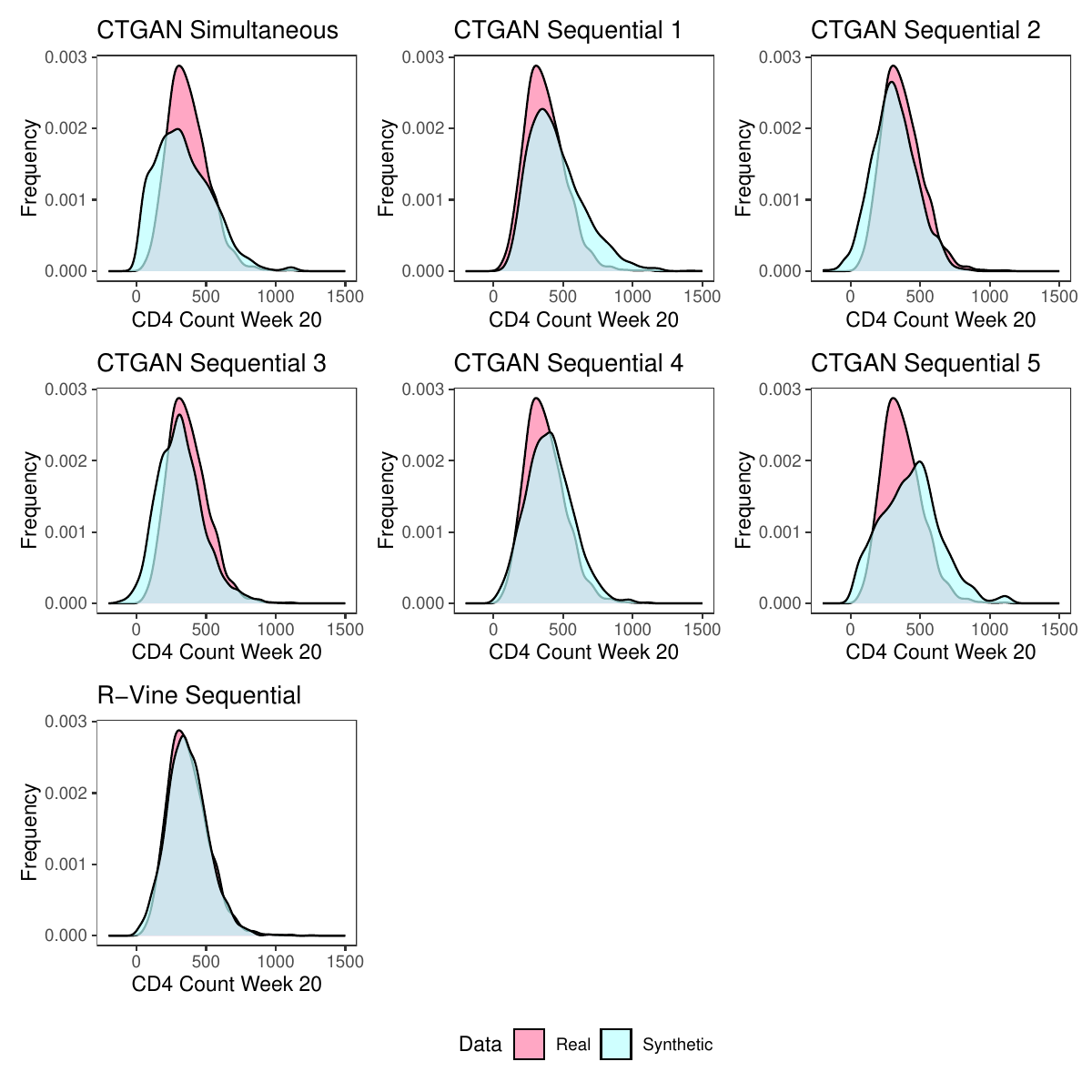}%
            \label{subfig:b}%
        }\\
        \subfloat[Run 3]{%
            \includegraphics[width=.45\linewidth]{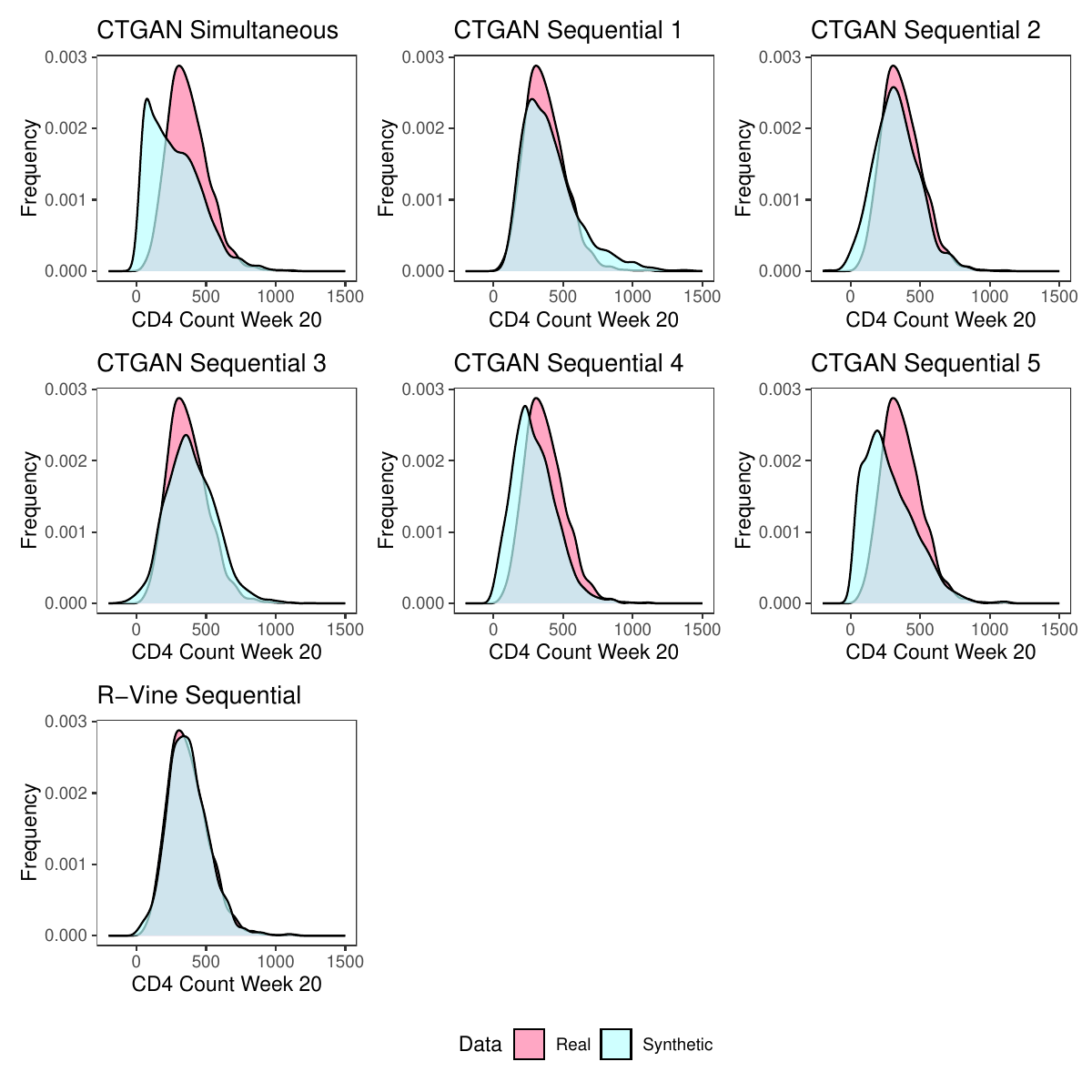}%
            \label{subfig:c}%
        }\hfill
        \subfloat[Run 4]{%
            \includegraphics[width=.45\linewidth]{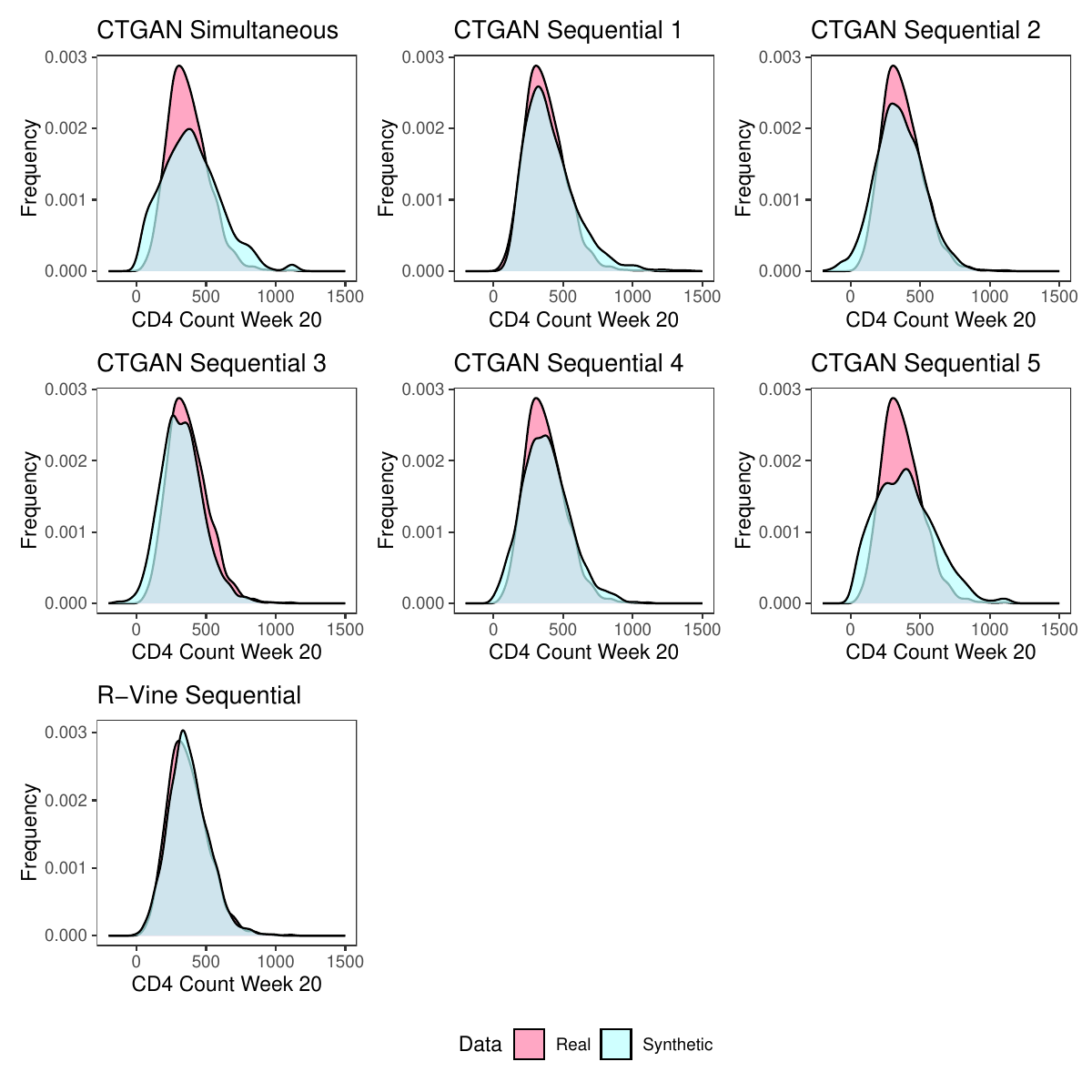}%
            \label{subfig:d}%
        }
        \caption{Density plots (pink for real data, blue for synthetic data) of CD4 cell count at week 20 across four different data generation runs for seven candidate synthetic data generators.}
        \label{fig:cd4204runs}
\end{figure*}

\label{applastpage}

\begin{figure}
    \centering
    \includegraphics[width=\linewidth]{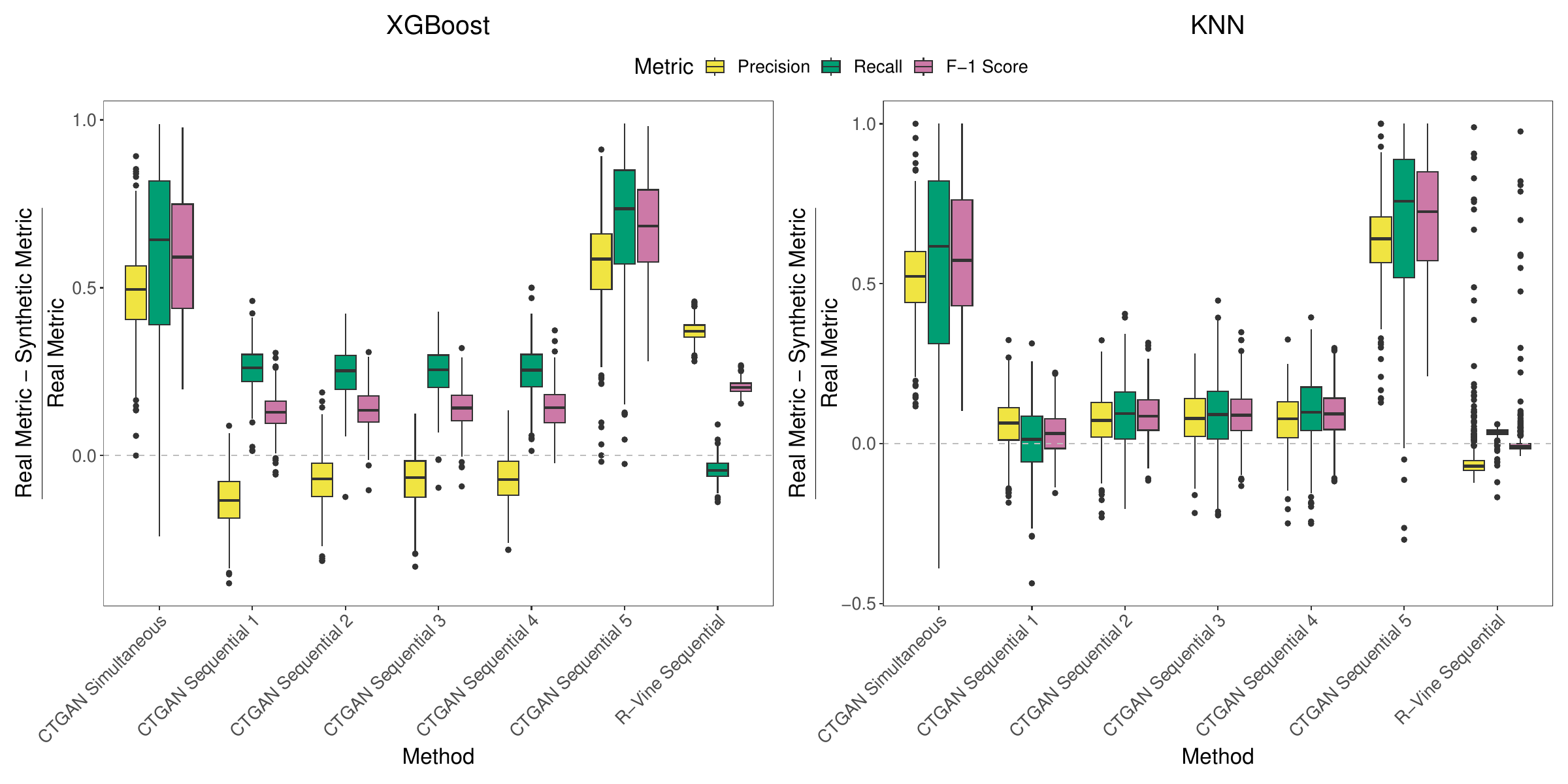}
    \caption{Comparison of the seven data generation methods by similarity of the ML efficacy metrics (precision, recall, F-1 score) for both XGBoost classifiers and k-nearest neighbors (KNN) classifiers. The vertical axis represents the relative difference between the real and synthetic metric values, $\frac{\text{Real}-\text{Synthetic}}{\text{Real}}$. The horizontal dashed line at zero indicates that the metric value for the classifier trained on real data is equal to that of the classifier trained on the synthetic data. A value close to zero indicates that the two models performed similarly and therefore the synthetic data successfully recreated the original data distribution.}
    \label{fig:7methods_tx_age}
\end{figure}

\end{document}